\documentclass[twocolumn]{aastex631}

\usepackage{}
\usepackage{gensymb}  
\usepackage{multirow}  

\shorttitle{GLIMPSE: direct MZR}
\shortauthors{Hsiao et al.}

\begin{document}

\title{A Glimpse of the Low-Mass End of the Direct Mass-Metallicity Relation at $z\sim6-8$}

\correspondingauthor{Tiger Hsiao}
\email{tiger.hsiao@utexas.edu}

\newcommand{\CfA}{\affiliation{Center for Astrophysics \text{\textbar} Harvard \& Smithsonian, 60 Garden Street, Cambridge, MA 02138, USA}}

\newcommand{\STScI}{\affiliation{Space Telescope Science Institute (STScI), 3700 San Martin Drive, Baltimore, MD 21218, USA}}

\newcommand{\JHU}{\affiliation{Center for Astrophysical Sciences, Department of Physics and Astronomy, The Johns Hopkins University, 3400 N Charles St. Baltimore, MD 21218, USA}}

\newcommand{\ESAAURA}{\affiliation{Association of Universities for Research in Astronomy (AURA), Inc.~for the European Space Agency (ESA)}}

\newcommand{\Austin}{\affiliation{Department of Astronomy, The University of Texas at Austin, 2515 Speedway, Austin, Texas 78712, USA}}

\newcommand{\BGU}{\affiliation{Physics Department, Ben-Gurion University of the Negev, P.O. Box 653, Be'er-Sheva 84105, Israel}}

\newcommand{\kapteyn}{\affiliation{Kapteyn Astronomical Institute, University of Groningen, 9700 AV Groningen, The Netherlands}}

\newcommand{\CAB}{\affiliation{Centro de Astrobiología (CAB), CSIC-INTA, Ctra. de Ajalvir km 4, Torrejón de Ardoz, E-28850, Madrid, Spain}}

\newcommand{\Cambridge}{\affiliation{Kavli Institute for Cosmology, University of Cambridge, Madingley Road, Cambridge CB3 0HA, UK}}

\newcommand{\Cavendish}{\affiliation{Cavendish Laboratory, University of Cambridge, 19 JJ Thomson Avenue, Cambridge CB3 0HE, UK}}

\newcommand{\UCL}{\affiliation{Department of Physics and Astronomy, University College London, Gower Street, London WC1E 6BT, UK}}

\newcommand{\Arizona}{\affiliation{Department of Astronomy / Steward Observatory, University of Arizona, 933 N Cherry Ave, Tucson, AZ 85721}}

\newcommand{\TsingHua}{\affiliation{Department of Astronomy, Tsinghua University, Beijing 100084, China}}

\newcommand{\Chiba}{\affiliation{Center for Frontier Science, Chiba University, 1-33 Yayoi-cho, Inage-ku, Chiba 263-8522, Japan}}

\newcommand{\ICRR}{\affiliation{Institute for Cosmic Ray Research, The University of Tokyo, 5-1-5 Kashiwanoha, Kashiwa, Chiba 277-8582, Japan}}

\newcommand{\NAOJ}{\affiliation{National Astronomical Observatory of Japan, 2-21-1 Osawa, Mitaka, Tokyo 181-8588, Japan}}

\newcommand{\Sokendai}{\affiliation{Department of Astronomical Science, SOKENDAI (The Graduate University for Advanced Studies), Osawa 2-21-1, Mitaka, Tokyo, 181-8588, Japan}}

\newcommand{\IPMU}{\affiliation{Kavli Institute for the Physics and Mathematics of the Universe (WPI), University of Tokyo, Kashiwa, Chiba 277-8583, Japan}}

\newcommand{\Dawn}{\affiliation{Cosmic Dawn Center (DAWN), Denmark}}

\newcommand{\Cop}{\affiliation{Niels Bohr Institute, University of Copenhagen, Jagtvej 128, DK-2200 Copenhagen N, Denmark}}

\newcommand{\Tokyo}{\affiliation{Department of Astronomy, Graduate School of Science, the University of Tokyo, 7-3-1 Hongo, Bunkyo, Tokyo 113-0033, Japan}}

\newcommand{\Oxford}{\affiliation{Department of Physics, University of Oxford, Denys Wilkinson Building, Keble Road, Oxford OX1 3RH, UK}}

\newcommand{\MIT}{\affiliation{MIT Kavli Institute for Astrophysics and Space Research, 70 Vassar Street, Cambridge, MA 02139, USA}}

\newcommand{\Dunlap}{\affiliation{David A. Dunlap Department of Astronomy and Astrophysics, University of Toronto, 50 St. George Street, Toronto, Ontario, M5S 3H4, Canada}}

\newcommand{\IAP}{\affiliation{Institut d'Astrophysique de Paris, CNRS, Sorbonne Université, 98bis Boulevard Arago, 75014, Paris, France}}

\newcommand{\CFC}{\affiliation{Cosmic Frontier Center, The University of Texas at Austin, Austin, TX 78712}}

\newcommand{\Rutgers}{\affiliation{Rutgers University, Department of Physics and Astronomy, 136 Frelinghuysen Road, Piscataway, NJ 08854, USA}}

\newcommand{\YaleAstro}{\affiliation{Department of Astronomy, Yale University, New Haven, CT 06511, USA}}

\newcommand{\YalePhys}{\affiliation{Department of Physics, Yale University, New Haven, CT 06511, USA}}
\newcommand{\BHI}{\affiliation{Black Hole Initiative, Harvard University, 20 Garden Street, Cambridge, MA 02138, USA}}

\newcommand{\unige}{\affiliation{Department of Astronomy, University of Geneva, Chemin Pegasi 51, 1290 Versoix, Switzerland}}

\newcommand{\Lyon}{\affiliation{Univ Lyon, Univ Lyon1, Ens de Lyon, CNRS, CRAL UMR5574, F-69230, Saint-Genis-Laval, France}}


\author[0000-0003-4512-8705]{Tiger Yu-Yang Hsiao} \Austin \CFC
\author[0000-0002-0302-2577]{John Chisholm} \Austin \CFC
\author[0000-0002-4153-053X]{Danielle A. Berg} \Austin \CFC
\author[0000-0001-8519-1130]{Steven L. Finkelstein} \Austin \CFC
\author[0000-0002-5588-9156]{Vasily Kokorev}
\Austin \CFC
\author[0000-0002-7570-0824]{Hakim Atek} \IAP
\author[0000-0003-3997-5705]{Rohan P.~Naidu}\altaffiliation{Hubble Fellow} \MIT
\author[0000-0001-7201-5066]{Seiji Fujimoto} \Dunlap
\author[0000-0001-6278-032X]{Lukas J. Furtak} \Austin \CFC

\author[0000-0002-8192-8091]{Angela Adamo}
\affiliation{Department of Astronomy, The Oskar Klein Centre, Stockholm University, AlbaNova, SE-10691 Stockholm, Sweden} 
\author[0000-0001-7578-2412]{Archana Aravindan} \Austin \CFC
\author[0000-0003-3983-5438]{Yoshihisa Asada}
\affiliation{Dunlap Institute for Astronomy and Astrophysics, 50 St. George Street, Toronto, Ontario M5S 3H4, Canada}
\author[0000-0001-8104-9751]{Arghyadeep Basu} \Lyon
\author[0000-0003-1609-7911]{J{\'e}r{\'e}my Blaizot} \Lyon
\author[0000-0002-7973-5442,sname=Choustikov,gname=Nicholas]{Nicholas Choustikov}
\Oxford
\author[0000-0002-5588-9156,sname=Dessauges-Zavadsky,gname=Miroslava]{Miroslava Dessauges-Zavadsky}
\unige
\author[0000-0001-7232-5355]{Qinyue Fei} \Dunlap
\author[0000-0003-1561-3814]{Harley Katz}
\affiliation{Department of Astronomy \& Astrophysics, University of Chicago, 5640 S Ellis Avenue, Chicago, IL 60637, USA}
\affiliation{Kavli Institute for Cosmological Physics, University of Chicago, Chicago IL 60637, USA}
\author[0000-0002-3897-6856]{Damien Korber} \unige
\author[0000-0001-5538-2614]{Kristen.~B.~W. McQuinn} \STScI \Rutgers
\author[0000-0002-3706-9955]{Marcie Mun} \IAP
\author[0000-0002-8984-0465]{Julian B. Mu\~noz} \Austin \CFC
\author[0000-0002-5554-8896]{Priyamvada Natarajan} \YaleAstro \YalePhys \BHI
\author[0000-0003-4717-0376]{Mabel G. Stephenson} \Austin \CFC
\author[0000-0001-7144-7182]{Daniel Schaerer} \unige





\newcommand{\LCDM}{$\Lambda$CDM}

\newcommand{\red}[1]{{\color{red} #1}}
\newcommand{\redss}[1]{{\color{red} ** #1}}
\newcommand{\redbf}[1]{{\color{red}\bf #1 \color{black}}}

\newcommand{\ny}{$\tilde {\rm n}$}
\newcommand{\about}{$\sim$}
\newcommand{\appr}{$\approx$}
\newcommand{\gt}{$>$}
\newcommand{\um}{$\mu$m}
\newcommand{\uJy}{$\mu$Jy}
\newcommand{\sig}{$\sigma$}
\newcommand{\Lya}{Lyman-$\alpha$}
\renewcommand{\th}{$^{\rm th}$}
\newcommand{\lam}{$\lambda$}

\newcommand{\tentothe}[1]{$10^{#1}$}
\newcommand{\tentotheminus}[1]{$10^{-#1}$}
\newcommand{\e}[1]{$\times 10^{#1}$}
\newcommand{\en}[1]{$\times 10^{-#1}$}
\newcommand{\cgsfluxunits}{erg$\,$s$^{-1}\,$cm$^{-2}$}
\newcommand{\linefluxunits}{\tentotheminus{20} \cgsfluxunits}

\newcommand{\logU}{$\log(U)$}
\newcommand{\logOH}{12+log(O/H)}

\newcommand{\sinv}{s$^{-1}$}
\newcommand{\kms}{km\,s$^{-1}$}

\newcommand{\footnoteurl}[1]{\footnote{\url{#1}}}

\newcommand{\tnm}[1]{\tablenotemark{#1}}
\newcommand{\super}[1]{$^{\rm #1}$}
\newcommand{\supa}{$^{\rm a}$}
\newcommand{\supb}{$^{\rm b}$}
\newcommand{\supc}{$^{\rm c}$}
\newcommand{\supd}{$^{\rm d}$}
\newcommand{\supe}{$^{\rm e}$}
\newcommand{\supf}{$^{\rm f}$}
\newcommand{\supg}{$^{\rm g}$}
\newcommand{\suph}{$^{\rm h}$}
\newcommand{\supi}{$^{\rm i}$}
\newcommand{\supj}{$^{\rm j}$}
\newcommand{\supk}{$^{\rm k}$}
\newcommand{\supl}{$^{\rm l}$}
\newcommand{\supm}{$^{\rm m}$}
\newcommand{\supn}{$^{\rm n}$}
\newcommand{\supo}{$^{\rm o}$}

\newcommand{\squared}{$^2$}
\newcommand{\cubed}{$^3$}

\newcommand{\sqarcmin}{arcmin\squared}

\newcommand{\supcomma}{$^{\rm ,}$}

\newcommand{\rhalf}{$r_{1/2}$}

\newcommand{\chisq}{$\chi^2$}

\newcommand{\Zgas}{$Z_{\rm gas}$}  
\newcommand{\Zstar}{$Z_*$}  

\newcommand{\per}{$^{-1}$}
\newcommand{\inv}{\per}
\newcommand{\Mstar}{$M^*$}
\newcommand{\Lstar}{$L^*$}
\newcommand{\phistar}{$\phi^*$}

\newcommand{\logM}{log($M_*$/\Msun)}

\newcommand{\LUV}{$L_{UV}$}
\newcommand{\MUV}{$M_{UV}$}

\newcommand{\Msun}{$M_\odot$}
\newcommand{\Lsun}{$L_\odot$}
\newcommand{\Zsun}{$Z_\odot$}

\newcommand{\Mvir}{$M_{vir}$}
\newcommand{\Mt}{$M_{200}$}
\newcommand{\Mf}{$M_{500}$}

\newcommand{\Ndotion}{$\dot{N}_{\rm ion}$}
\newcommand{\xiion}{$\xi_{\rm ion}$}
\newcommand{\logxiion}{log(\xiion)}
\newcommand{\fesc}{$f_{\rm esc}$}

\newcommand{\XHI}{$X_{\rm HI}$}
\newcommand{\XHII}{$X_{\rm HII}$}
\newcommand{\RHII}{$R_{\rm HII}$}

\newcommand{\Halpha}{H$\alpha$}
\newcommand{\Hbeta}{H$\beta$}
\newcommand{\Hgamma}{H$\gamma$}
\newcommand{\Hdelta}{H$\delta$}
\newcommand{\Halphaw}{\Halpha\,$\lambda$6563}
\newcommand{\Hbetaw}{\Hbeta\,$\lambda$4861}
\newcommand{\Hgammaw}{H$\gamma$\,$\lambda$4340}
\newcommand{\Hdeltaw}{H$\delta$\,$\lambda$4101}
\newcommand{\Ha}{\Halpha}
\newcommand{\Hb}{\Hbeta}

\newcommand{\I}{\,{\sc i}}
\newcommand{\II}{\,{\sc ii}}
\newcommand{\III}{\,{\sc iii}}
\newcommand{\IV}{\,{\sc iv}}
\newcommand{\V}{\,{\sc v}}
\newcommand{\VI}{\,{\sc vi}}
\newcommand{\VII}{\,{\sc vii}}
\newcommand{\VIII}{\,{\sc viii}}

\newcommand{\HI}{H\I}
\newcommand{\HII}{H\II}
\newcommand{\HeI}{He\I}
\newcommand{\HeII}{He\II}

\newcommand{\CII}{[C\II]}
\newcommand{\CIIw}{\CII\,$\lambda$2325 (blend)}
\newcommand{\CIII}{[C\III]}
\newcommand{\CIIIw}{\CIII\,$\lambda$1908}
\newcommand{\CIIId}{C\III]}
\newcommand{\CIIIdw}{C\III]\,$\lambda\lambda$1907,1909}
\newcommand{\CIV}{C\IV}
\newcommand{\CIVw}{\CIV\,$\lambda$1549}
\newcommand{\OII}{[O\II]}
\newcommand{\OIIw}{\OII\,$\lambda$3727}
\newcommand{\OIIdw}{\OII\,$\lambda\lambda$3727,3730}
\newcommand{\OIII}{[O\III]}
\newcommand{\OIIIw}{\OIII\,$\lambda$5008}
\newcommand{\OIIIww}{\OIII\,$\lambda\lambda$4960,5008}
\newcommand{\OIIIdw}{\OIIIww}
\newcommand{\OIIIwa}{\OIII\,$\lambda$4364}
\newcommand{\OIIIwb}{O\III]\,$\lambda$1666}
\newcommand{\OIIIwc}{\OIII\,$\lambda$4960}
\newcommand{\NeIII}{[Ne\III]}
\newcommand{\NeIIIw}{\NeIII\,$\lambda$3870}
\newcommand{\NeIIIwb}{\NeIII\,$\lambda$3969}
\newcommand{\NII}{[N\II]}
\newcommand{\NIIw}{\NII\,$\lambda$6585}
\newcommand{\NIIww}{\NII\,$\lambda$6550,$\lambda$6585}
\newcommand{\SII}{[S\II]}
\newcommand{\SIIdw}{\SII\,$\lambda$6718,$\lambda$6733}
\newcommand{\HeIw}{\HeI\,$\lambda$3889}
\newcommand{\HeIwa}{\HeI\,$\lambda$4473}
\newcommand{\HeIIw}{\HeII\,$\lambda$1640}
\newcommand{\HeIIwb}{\HeII\,$\lambda$4687}
\newcommand{\NIII}{N\III]}
\newcommand{\NIV}{N\IV]}
\newcommand{\NIIIw}{\NIII\,$\lambda$1748}
\newcommand{\NIVw}{\NIV\,$\lambda$1486}
\newcommand{\MgII}{Mg\II}
\newcommand{\MgIIw}{\MgII\,$\lambda$2800}

\newcommand{\Lyaw}{Ly$\alpha$\,$\lambda$1216}



\newcommand{\Om}{\Omega_{\rm M}}
\newcommand{\OL}{\Omega_\Lambda}

\newcommand{\etal}{et al.}

\newcommand{\citeps}{\citep}

\newcommand{\HST}{{\em HST}}
\newcommand{\SST}{{\em SST}}
\newcommand{\Hubble}{{\em Hubble}}
\newcommand{\Spitzer}{{\em Spitzer}}
\newcommand{\Chandra}{{\em Chandra}}
\newcommand{\JWST}{{\em JWST}}
\newcommand{\Planck}{{\em Planck}}

\newcommand{\Bradac}{{Brada\v{c}}}

\newcommand{\citepeg}[1]{\citep[e.g.,][]{#1}}

\newcommand{\range}[2]{\! \left[ _{#1} ^{#2} \right] \!}  

\newcommand{\grizli}{\textsc{grizli}}
\newcommand{\eazypy}{\textsc{eazypy}}
\newcommand{\msaexp}{\textsc{msaexp}}
\newcommand{\trilogy}{\textsc{trilogy}}
\newcommand{\bagpipes}{\textsc{bagpipes}}
\newcommand{\beagle}{\textsc{beagle}}
\newcommand{\photutils}{\textsc{photutils}}
\newcommand{\SEP}{\textsc{sep}}
\newcommand{\piXedfit}{\textsc{piXedfit}}
\newcommand{\pyneb}{\textsc{pyneb}}
\newcommand{\HIIC}{\textsc{hii-chi-mistry}}
\newcommand{\astropy}{\textsc{astropy}}
\newcommand{\astrodrizzle}{\textsc{astrodrizzle}}
\newcommand{\multinest}{\textsc{multinest}}
\newcommand{\cloudy}{\textsc{Cloudy}}
\newcommand{\jdaviz}{\textsc{Jdaviz}}

\renewcommand{\tt}[1]{\texttt{#1}}

\newcommand{\SE}{\tt{SourceExtractor}}

\newcommand{\PD}[1]{\textcolor{blue}{[PD: #1\;]}}


\begin{abstract}
The competition between metal synthesis and feedback from massive stars establishes the mass-metallicity relation (MZR) at low-redshifts.
Examining this relation at higher redshifts, particularly at the low-mass end $\lesssim10^{8}\,{\rm M_\odot}$, is essential for understanding chemical enrichment and stellar feedback.
In this study, we utilize the deep ($\sim30\,$hrs) JWST/NIRSpec G395M GLIMPSE-D survey of the lensed field Abell S1063, to explore the low-mass end of the MZR at high redshift ($z\sim6-8$).
We identify eight \OIIIwa\ emitters, enabling the most reliable ``direct'' metallicity measurements in galaxies down to stellar masses of $\sim10^{6-8}\,{\rm M_\odot}$.
By combining our sample and galaxies with \OIIIwa\ detections from the literature, we calculate direct metallicities for 21 galaxies.
We compare our direct metallicities to those derived from strong-line diagnostics, and find them to be consistent with previous calibrations.
We fit the MZR at $10^{6.7-9}\,M_{\odot}$ 
with $\sim0.3-0.5$ dex lower metallicity than local galaxies at similar stellar mass.
We find the slope to be $0.25\pm0.10$, comparable to the local MZR; and the MZR exhibits a scatter of $\sim0.2\,$dex, which is larger than the local MZR, 
The lower metallicities may reflect denser, more gas-rich early environments, with continuous inflow of metal-poor gas diluting the ISM metallicity.
In addition, we show that in extremely high electron densities ($n_e \gtrsim 10^5\,{\rm cm^{-3}}$), metallicities can be significantly underestimated ($\sim0.5$ dex), if lower $n_e$ are assumed for galaxies with high $n_e$. 
In a nutshell, these observations provide the first glimpse of the low-mass MZR at $z\sim6-8$ using direct metallicity measurements.
More deep spectroscopic observations in lensed fields will be critical to robustly characterize the MZR and chemical evolution in the early universe.
\end{abstract}


\section{Introduction} \label{sec:intro}
Pristine gas flows into galaxies, and some fraction cools and is converted into stars.
Stars synthesize metals and release them back into the gas.
Some fraction of the stars inject energy and momentum into the interstellar medium (ISM), driving metal-enriched gas out of galaxies. More massive galaxies create more metals, and have deeper gravitational potentials, making it harder for star formation feedback to remove the gas.
This means low-mass galaxies are not massive enough to retain as much of the synthesized metals, leading to the ``Mass-Metallicity Relation'' (MZR), first identified in \citet{Lequeux1979}.

With the unprecedented data from the Sloan Digital Sky Survey \citep{York2000}, \citet{Tremonti2004} analyzed $\sim53,400$ star-forming galaxies and revealed a remarkably tight ($\pm0.05\,$dex) correlation between stellar mass and metallicity at $z\sim0$ in  massive galaxies ($\sim10^{9-11}\,M_{\odot}$) where galaxies with larger stellar mass have higher gas-phase metallicities. 
Subsequent works extended the MZR to lower-mass galaxies ($\sim10^{6}\,{\rm M_\odot}$; \citealt{Lee2006,Berg2012}).
More recent studies have traced the MZR to earlier epochs, identifying it at the onset of Cosmic Noon \citep[$z\sim3.5$; e.g.,][]{Savaglio2005,Erb2006,Maiolino2008,Mannucci2009,Zahid2011,Zahid2014,Zahid2014b,Steidel2014,Sanders2015,Sanders2018,Sanders2020,Sanders2021,He2024,Khostovan2025}.
These studies have shown that the overall metallicities decrease at fixed stellar mass with increasing redshift, indicating that MZR does evolve.
For instance, at $M_{*}\sim10^{10}\,M_{\odot}$, $z\sim2$ and $z\sim3$ galaxies exhibit metallicities $\sim0.3$ and $\sim0.4$ dex fewer metals than local galaxies, respectively.
However, in addition to the pure relation between metallicity and stellar mass, in the local universe, galaxies with higher star formation rates (SFRs) appear to be more metal-poor at fixed stellar mass \citep{Ellison2008,Mannucci2010,Curti2020,Sanders2021}, since at fixed stellar mass, higher SFR drives stronger outflows that expel metal-enriched gas, resulting in lower gas-phase metallicity.
This SFR-MZ relation holds up to $z\sim3.3$, indicating that the interplay between stellar mass, metal content, and star formation was already established when the Universe was only about 2 billion years old. This relation was thus referred as Fundamental Metallicity Relation (FMR).
The FMR naturally explains the observed evolution of the MZR from $z\sim3$ to $z=0$ as a projection effect from SFR.

To measure metallicity, the most reliable and precise approach is the so-called ``direct'' electron temperature ($T_{e}$) method \citep[e.g.,][]{Peimbert1967,Peimbert1969,Aller1984,Osterbrock1989,Izotov2006}.
In short, the direct method requires two transitions from the same ion that have excited levels with different energies. 
For instance, the \OIIIwa\ auroral line is a transition from $^1{\rm S}_{0}$, which has a higher excitation energy of $5.35\,{\rm eV}$ compared to \OIIIw\ ($^1{\rm D}_{2}$ $\rightarrow$ $^3{\rm P}_{2}$) with an excitation energy of $2.51\,{\rm eV}$.
Their excitation rates depend strongly on $T_{e}$ through the Boltzmann distribution.
Once $T_{e}$ and electron density are measured, the emissivity of each line can be calculated, the ionic abundance then comes from comparing the emissivities of O$^{++}$ to the hydrogen lines, which can be added to determine total abundances.

However, auroral lines require significantly higher excitation energies and are therefore much fainter than strong lines such as \OIIIw\ (under typical temperatures of $\sim10,000\,$K of HII regions).
Moreover, before JWST, auroral lines at $z\gtrsim1$ that were redshifted into the NIR were extremely arduous to detect since there was no space based observatory to make the measurements and galaxies are even fainter.
As a result, metallicity estimates often rely on so-called ``strong-line'' diagnostics instead \citep[e.g.,][]{Kewley2002,Pettini2004,Nagao2006,Kewley2008,Marino2013,Curti2017,Curti2020}, which are empirical calibrations that relate observed ratios of bright emission lines such as \OIIIw\ and H$\beta$ to the metallicity.
However, adopting strong-line calibrations anchored at $z\sim0$ with varying H$\beta$ equivalent widths  \citep{Nakajima2022} have been shown to be independent of redshift \citep{Langeroodi2026}.
These strong-line diagnostics have been calibrated using samples where the metallicity has been measured with the direct method, enabling metallicity measurements for much larger sample sizes, especially for faint or low-mass galaxies.


JWST stands as one of humanity’s greatest wonders thanks to its exceptional sensitivity and unique infrared wavelength coverage \citep{Gardner2006,Rigby2023,Gardner2023}, which provides an unprecedented opportunity to explore the evolution of the MZR and FMR at higher redshifts ($z\gtrsim3.5$). 
After three years of operation, astronomers have obtained the first glimpses of the MZR in the Epoch of Reionization (EoR; $z \gtrsim 6$) 
\citep[e.g.,][]{Heintz2023,Curti2024,Morishita2024,Chemerynska2024,Arellano2025,Sarkar2025,Chakraborty2025,Hsiao2025,Nishigaki2025,Pollock2026,Koller2026}.
These works have found a variety of slopes and normalizations that do not match predictions from simulations \citep[e.g.,][]{Torrey2019,Langan2020,Ucci2023,Marszewski2024}.
Besides, unlike the global FMR established across $z\sim0-3$, galaxies at $z>3$ are less enriched than local FMR predictions for a given stellar mass and SFR \citep{Heintz2023,Nakajima2023,Curti2024}, which might indicate that our understanding of the regulation of star-forming activity, stellar and metal content is incomplete at higher redshift.

Auroral lines (e.g., \OIIIwb\ and \OIIIwa) have now been detected out to $z \sim 11$--12, near the onset of the EoR \citep{Bunker2023,Hsiao2024a,Castellano2024}, enabling the direct metallicity measurements with \OIIIw\ detections with MIRI \citep{Hsiao2024b,Alvarez2025}.
Therefore, JWST has now opened a new window to study and calibrate the strong-line relations at higher redshift.
Early works in the first few years of JWST provided the first glimpses into strong-line diagnostics \citep[e.g.,][]{Nakajima2023,Laseter2024,Sanders2024,Scholte2025,Chakraborty2025,Sanders2025}.
These studies usually focus on a wide redshift range, spanning, for example, $z\sim2-10$, to ensure large statistical samples, and are usually limited to galaxies with $M_{*}\gtrsim10^{8}\,M_{\odot}$.
As a result, most high-redshift MZR studies in the lower mass regime still relied on strong-line diagnostics, which remain poorly calibrated \citep[e.g.,][]{Chemerynska2024,Hsiao2025,Asada2026}.
Consequently, the low-mass end of the MZR at high redshift remains in an early stage, and direct metallicity measurements for low-mass galaxies are needed.
It is, therefore, crucial to test whether strong-line calibrations can reliably diagnose trends in high-z populations.

To investigate the low-mass end of the high-redshift ($z>6$) MZR and FMR, we utilize data from the JWST Cycle 2 deep NIRCam imaging survey GLIMPSE (GO 3293; PI Atek $\&$ Chisholm) and the JWST NIRSpec spectroscopy survey GLIMPSE-D program (DDT 9223; PI Fujimoto $\&$ Naidu).
GLIMPSE has produced JWST’s deepest NIRCam imaging dataset (after accounting for lensing magnification; \citealt{Atek2025}). 
The GLIMPSE-D program simultaneously obtained 30 hours of G395M spectroscopy for additional galaxies through the micro-shutter assembly (MSA), providing the best dataset to date for probing the faint end of the MZR at high redshift, where the auroral line \OIIIwa\ and \OIIIw\ are accessible in G395M for $5.5 \lesssim z \lesssim 9.2$.

In this work, we report the discovery of 8 star-forming galaxies with \OIIIwa\ detections, for which we measure gas-phase metallicities via the direct method and determine their stellar masses from spectral energy distribution (SED) fitting.
This enables us to probe the high-redshift MZR down to the low-mass regime (${\rm log} (M/M_{\odot})<8$) using direct-method metallicity measurements.
The paper is organized as follows.
In \S\ref{Sec:odm}, we describe the observations and the data utilized in this paper, and the emission line measurements are presented in \S\ref{sec:fitting}. 
Then, we present the analysis of the MZR and FMR in \S\ref{Sec:analysis}.
We discuss the high-z MZR in \S\ref{Sec:discussion} and conclude in \S\ref{sec:conclusion}.

Throughout this article, we adopt solar abundance ratios
\logOH\ = 8.69 \citep{Asplund2021}.
Where needed, we adopt the {\em Planck} 2018 flat \LCDM\ cosmology \citep{Planck18_cosmo}
with $H_0 = 67.7$ km s\inv\ Mpc\inv, $\Om = 0.31$, and $\OL = 0.69$.

\section{Observations and Data Reduction}
\label{Sec:odm}

This paper primarily makes use of the NIRCam photometric data from JWST Cycle 2 deep survey GLIMPSE \citep[GO 3293; PI Atek $\&$ Chisholm;][]{Atek2025} and the follow-up JWST GLIMPSE-D program (DDT 9223; PI Fujimoto $\&$ Naidu), targeting one of the Frontier Fields lensing cluster: Abell S1063 at $z=0.348$.
We utilize GLIMPSE-D NIRSpec spectroscopic data to measure the line fluxes of emission lines and therefore the metallicity as well as the star-formation rate.
GLIMPSE NIRCam photometric data is used for SED fitting, to determine the stellar mass and star-formation rate.

\subsection{GLIMPSE-D NIRSpec Spectroscopy}
\label{sec:spec}

GLIMPSE-D NIRSpec spectroscopy was obtained to confirm the nature of the Population III candidate GLIMPSE-16043, discovered in the GLIMPSE survey \citep{Fujimoto2025}.
GLIMPSE-D conducted the G395M/F290LP grating spectroscopy ($R\sim1000$) with a total exposure time of 29.78 hours, making it one of the deepest NIRSpec spectroscopy surveys, suitable for the study faint/low-mass galaxies. 
A standard 3-point nod pattern is executed for each MSA configuration, with the NRSIRS2 readout mode.
Using the Multi-Object Spectroscopy (MOS) mode, GLIMPSE-D obtained 384 spectra, and the observations were conducted between June 30th and July 2nd, 2025.
The sources are selected, from the main footprint based on the primary target and then included as many phot-z candidates with redshifts between 4-8 with observed  magnitude brighter than 28th mag.
We used multiple SED fitting routines and a combination of photo-zs and Lyman breaks to compile the full sample.

To ensure the consistency and reduce systematic errors across different datasets, both GLIMPSE-D and literature spectra (see \S\ref{sec:archival}) are reduced, measured, and an analyzed in a uniform way.
GLIMPSE-D MSA spectra were reduced using \texttt{msaexp} \citep{Brammer2023}.
Literature data analyzed in this paper were obtained from the DAWN JWST Archive\footnote{\url{https://dawn-cph.github.io/dja/}} (DJA) was also uniformly-reduced by \texttt{msaexp} pipeline.
More details of the data reduction can be found in \citet{Kokorev2025b}.

\subsection{GLIMPSE NIRCam and Ancillary Photometry}
\label{sec:phot}

GLIMPSE obtained JWST/NIRCam imaging in seven broadband filters: F090W, F115W, F200W, F277W, F356W and F444W, and two medium-band filters: F410M and F480M.
GLIMPSE yields a 5-$\sigma$ depth of $\sim30.7$ ABmag (0\farcs2 diameter aperture), under $\sim$ 20--40 hours exposure time in each filter, making it one of the deepest JWST imaging programs to date, with a total of $\sim120$ hours of exposure \citep{Atek2025}.

In addition to JWST NIRCam photometry, we use ancillary photometry.
This includes Hubble Space Telescope (HST) ACS and WFC3 IR and UVIS imaging from the Hubble Frontier Fields \citep[HFF; PID 14037, PI Lotz;][]{Lotz2017}, BUFFALO \citep[PID 15117, PI Steinhardt;][]{Steinhardt2020}, and FLASHLIGHTS (PID 15936, PI Kelly) programs.

Again, to ensure a uniform reduction, all of these data above mentioned are reduced using JWST Science Calibration Pipeline (version \texttt{v12.0.9}), with the Calibration Reference Data System (CRDS) context file \texttt{jwst$\_$1321.pmap} hosted on DJA \citep{Endsley2023}.
We remove the artificial emission, such as the ``snowballs'', the ``wisps'', account for the diffraction spikes and correct for the 1/f noise.
The details of GLIMPSE imaging data can be found in \citet{Atek2025}, and the reduction of the literature data obtained from DJA is described in detail in \citet{Valentino2023}.

For the photometry of \OIIIwa\ emitters, we acquire the photometry again through DJA.
In short, photometry is measured within a circular aperture of D=0\farcs1 using \texttt{photutils} \citep{Bradley2020}, and uncertainties are estimated by placing random apertures in 2000 source-free regions near each galaxy.
Eight out of 13 of these data have JWST NIRCam imaging data, with at least 8 filters, enough to cover the rest-frame optical to determine their physical properties using SED fitting.

\subsection{Archival Data}
\label{sec:archival}

Besides GLIMPSE-D spectroscopic data, we collect literature data for 13 galaxies that has detected ($>3\sigma$) \OIIIwa\ at the redshift of interest, $z\sim6-8$, (see \S\ref{sec:sample}).
These data include: Early Release Observations \citep[ERO; GO 2736; PI: Pontoppidan;][]{Curti2023,Nakajima2023}, GTO 1199 \citep[PI: Stiavelli;][]{Morishita2024}, EXCELS \citep[GO 3543; PIs: Carnall and Cullen;][]{Scholte2025,Stanton2025}, and AURORA \citep[GO 1914, PIs: Shapley and Sanders;][]{Sanders2025}.
For the details of each observation, we encourage readers to explore the individual papers.
All these data are obtained through the DJA of version 4.4.
For the details of each JWST observation of the literature, including the exposure times and the mosaic design, we encourage the readers to check the individual papers, which can be found in Table \ref{tab:info}.

\section{Emission Line Measurements}
\label{sec:fitting}

\subsection{Line Fits}

We perform uniform fits to all spectra from GLIMPSE-D, and our literature sample using \texttt{msaexp$\_$OLF} \citep{Kokorev2024,Kokorev2025b}, which is an expanded line fitting code built on top of \texttt{msaexp} \citep{Brammer2023} with more flexible features.
In short, \texttt{msaexp$\_$OLF} adopts Gaussian profiles to model individual emission lines with splines to fit the continuum.
For the selection parameters, we allow the redshift to vary within $\pm0.05$ from the preliminary spectroscopic redshift determination from DJA/\texttt{msaexp}
, with the velocity (line width) of $50-850\,{\rm km/s}$, including 10 splines to fit the continuum.
The full spectrum is convolved with the NIRSpec line spread function.

These processes are repeated in both GLIMPSE-D and literature data.
If grating (medium-resolution and high-resolution) observations exist, since \OIIdw\ is a doublet, a fit with a single width would underestimate the line fluxes.
Thus, we allow a different width and we refit \OIIdw\ where applicable.
We note that one of the literature sample, COSMOS-419213, is missing in the DJA catalog.
Therefore, we do not fit the lines nor measure the metallicity in COSMOS-419213.
Measured emission lines from both GLIMPSE-D and literature sample including \OIIw, \NeIIIw, H$\gamma$, \OIIIwa, H$\beta$, \OIIIwc, \OIIIw, and H$\alpha$, are listed in Table \ref{tab:spec}.
\begin{deluxetable}{lcrrc}
\tablecaption{\label{tab:info}
The spectroscopic redshifts, RA and DEC, and the reference of GLIMPSED \OIIIwa\ emitters as well as the literature data.}
\tablewidth{\columnwidth}
\tablehead{
\colhead{ID} &
\colhead{$z_{\rm spec}$} &
\colhead{RA} &
\colhead{DEC} &
\colhead{Reference${^a}$} 
\\
\colhead{} &
\colhead{} &
\colhead{deg} &
\colhead{deg} &
\colhead{}
}
\startdata
\multicolumn{5}{c}{GLIMPSED} \\
\hline
13882 & $6.107$ & 342.23651 & -44.54689 & \ldots\\
45883 & $6.222$ & 342.24643 & -44.54088 & \ldots\\
6170 & $6.437$ & 342.25748 & -44.55910 & \ldots\\
12801 & $7.569$ & 342.24939 & -44.54817 & \ldots \\
5381 & $7.549$ & 342.26215 & -44.56067 &  \ldots \\
10357 & $6.060$ & 342.24915 & -44.55183 &  \ldots\\
2864 & $6.489$ & 342.23004 & -44.56598 &  \ldots\\
45572 & $7.318$ & 342.23804 & -44.54012 &  \ldots\\
\hline
\multicolumn{5}{c}{Literature} \\
\hline
GTO1199-2 & $7.228$ & 177.41771 & 22.41743 & 3 \\
COSMOS-419213 & $6.809$ & \ldots & \ldots & 6 \\
CEERS-01027 & $7.822$ & 214.88299 & 52.84042 & 2 \\
ERO-05144 & $6.380$ & 110.83967 & -73.44536 & 1, 2 \\
ERO-06355 & $7.665$ & 110.84459 & -73.43506 & 1, 2 \\
GLASS-10021 & $7.286$ & 3.60851 & -30.41854 & 2 \\
ERO-10612 & $7.659$ & 110.83396 & -73.43452 & 1, 2 \\
EXCELS-48659 & $6.794$ & 34.25147 & -5.24653 & 4\\
EXCELS-60713 & $6.169$ & 34.31350 & -5.22674 & 5 \\
GLASS-100003 & $7.878$ & 3.60451 & -30.38044 & 2 \\
GOODSN-100026 & $7.206$ & 189.22513 & 62.28629 & 6 \\
GOODSN-100163& $6.747$ & 189.13811 & 62.27444 & 6 \\
EXCELS-119504 & $7.916$ & 34.40249 & -5.13447 & 4\\
\enddata
\tablenotetext{a}{A reference is provided here to the work in which \OIIIwa\ was originally reported. All lines measured in this paper are the uniform fits performed in this work instead.
(1.) \citet{Curti2023},
(2.) \citet{Nakajima2023},
(3.) \citet{Morishita2024},
(4.) \citet{Scholte2025},
(5.) \citet{Stanton2025},
(6.) \citet{Sanders2025}.
}
\end{deluxetable}

\begin{deluxetable*}{lcccccccc}
\tablecaption{\label{tab:spec}
The line fluxes of emission lines of the $z\sim6-8$ Auroral Sample.}
\tablewidth{\columnwidth}
\tablehead{
\colhead{ID} &
\colhead{\OIIw} &
\colhead{\NeIIIw} &
\colhead{H$\gamma$} &
\colhead{\OIIIwa} &
\colhead{H$\beta$} &
\colhead{\OIIIwc} &
\colhead{\OIIIw} &
\colhead{H$\alpha$} 
}
\startdata
\multicolumn{9}{c}{GLIMPSED} \\
\hline
13882 & \ldots & \ldots & $ 44\pm  8$ & $ 21\pm  8$ & $109\pm  7$ & $230\pm  8$ & $648\pm 12$ & $373\pm 11$ \\
45883 & \ldots & \ldots & $ 27\pm  3$ & $ 11\pm  3$ & $ 55\pm  3$ & $ 97\pm  3$ & $343\pm  6$ & $156\pm  5$ \\
6170 & \ldots & $ 28\pm  6$ & $ 32\pm  4$ & $ 12\pm  3$ & $ 58\pm  3$ & $159\pm  5$ & $434\pm  7$ & $163\pm  6$ \\
12801 & $ 28\pm  3$ & $ 44\pm  3$ & $ 37\pm  3$ & $ 16\pm  2$ & $ 74\pm  3$ & $163\pm  4$ & $487\pm  7$ & \ldots \\
5381 & $ 94\pm  5$ & $ 65\pm  5$ & $ 51\pm  5$ & $ 14\pm  4$ & $110\pm  5$ & $249\pm  6$ & $784\pm 12$ & \ldots \\
10357 & \ldots & \ldots & $ 85\pm  6$ & $ 40\pm  5$ & $173\pm  6$ & $338\pm  7$ & $1116\pm 15$ & $405\pm  8$ \\
2864 & \ldots & $ 24\pm  6$ & $ 22\pm  4$ & $ 15\pm  3$ & $ 64\pm  3$ & $123\pm  4$ & $329\pm  6$ & $163\pm  6$ \\
45572 & \ldots & $ 37\pm  3$ & $ 61\pm  3$ & $ 19\pm  3$ & $ 85\pm  3$ & $160\pm  4$ & $473\pm  7$ & \ldots \\
\hline
\multicolumn{9}{c}{Literature} \\
\hline
GTO1199-2 & $329\pm  8$ & $310\pm  7$ & $285\pm  7$ & $154\pm  7$ & $642\pm  9$ & $1366\pm 15$ & $4097\pm 32$ & \ldots \\
COSMOS-419213 & \ldots & \ldots & \ldots & \ldots & \ldots & \ldots & \ldots & \ldots \\
CEERS-01027 & $ 44\pm  6$ & $ 78\pm  6$ & $ 73\pm  5$ & $ 26\pm  4$ & $144\pm  5$ & $340\pm  7$ & $1044\pm 12$ & \ldots \\
ERO-05144 & $331\pm 40$ & $112\pm  8$ & $116\pm  6$ & $ 32\pm  5$ & $218\pm  6$ & $476\pm  8$ & $1422\pm 15$ & $518\pm 12$ \\
ERO-06355 & $514\pm 13$ & $239\pm  8$ & $228\pm  8$ & $ 52\pm  6$ & $443\pm 10$ & $1202\pm 15$ & $3447\pm 29$ & \ldots \\
GLASS-10021 & $258\pm 37$ & $232\pm 25$ & $204\pm 19$ & $108\pm 20$ & $538\pm 26$ & $1165\pm 32$ & $3286\pm 56$ & \ldots \\
ERO-10612 & $ 43\pm  4$ & $ 75\pm  3$ & $ 67\pm  3$ & $ 28\pm  3$ & $124\pm  4$ & $295\pm  5$ & $907\pm 10$ & \ldots \\
EXCELS-48659 & \ldots & \ldots & $ 30\pm 10$ & $ 28\pm 10$ & $ 66\pm  8$ & $134\pm  9$ & $371\pm 12$ & $218\pm 15$ \\
EXCELS-60713 & $ 89\pm 11$ & $ 51\pm  9$ & $ 50\pm  5$ & $ 19\pm  9$ & $100\pm  5$ & $221\pm  6$ & $680\pm  9$ & $278\pm  9$ \\
GLASS-100003 & $ 48\pm 10$ & $ 61\pm 11$ & $ 83\pm 12$ & $ 38\pm 10$ & $165\pm 13$ & $406\pm 15$ & $1134\pm 19$ & \ldots \\
GOODSN-100026 & $ 80\pm  7$ & $ 73\pm  7$ & $ 59\pm  4$ & $ 22\pm  4$ & $108\pm  5$ & $245\pm  5$ & $711\pm  9$ & $205\pm 14$ \\
GOODSN-100163 & $ 53\pm  7$ & $ 34\pm  4$ & $ 33\pm  3$ & $ 13\pm  3$ & $ 66\pm  4$ & $192\pm  5$ & $533\pm  7$ & $210\pm  9$ \\
EXCELS-119504  & $ 45\pm  9$ & $ 50\pm  7$ & $ 44\pm  6$ & $ 21\pm  6$ & $ 88\pm  7$ & $219\pm 10$ & $617\pm 12$ & \ldots \\
\enddata
\tablenotetext{}{\textbf{Note.} All emission lines are shown in the unit of $10^{-20}\,{\rm erg\,s^{-1}\,cm^{-2}}$.
}
\end{deluxetable*}

\subsection{The $z\sim6-8$ Auroral Sample}
\label{sec:sample}
There are a total of 16 galaxies with robust \OIIIwa\ detection (SNR(\OIIIwa) $>3$) in GLIMPSE-D sample.
Next, we examine these \OIIIwa; 8 out of 16 galaxies have broader permitted emission lines, which might be indicative of active galactic nuclei (AGN), or other outflow processes.
The detailed AGN identification can be found in \citet{Fei2025}, which includes: 11026, 12248, 41299, 46938, 55241, and 41948 (a lensed image of 39803).
GLIMPSED-329380 also has broad permitted emission lines, which possibly comes from an AGN driven outflow.
This galaxy is excluded and analyzed in details in \citet{Korber2026}.
GLIMPSED-26653 is identified as a possible outflow broad-line emitter, which is also excluded from the sample.

\begin{figure*}
\centering
\includegraphics[width=\textwidth]{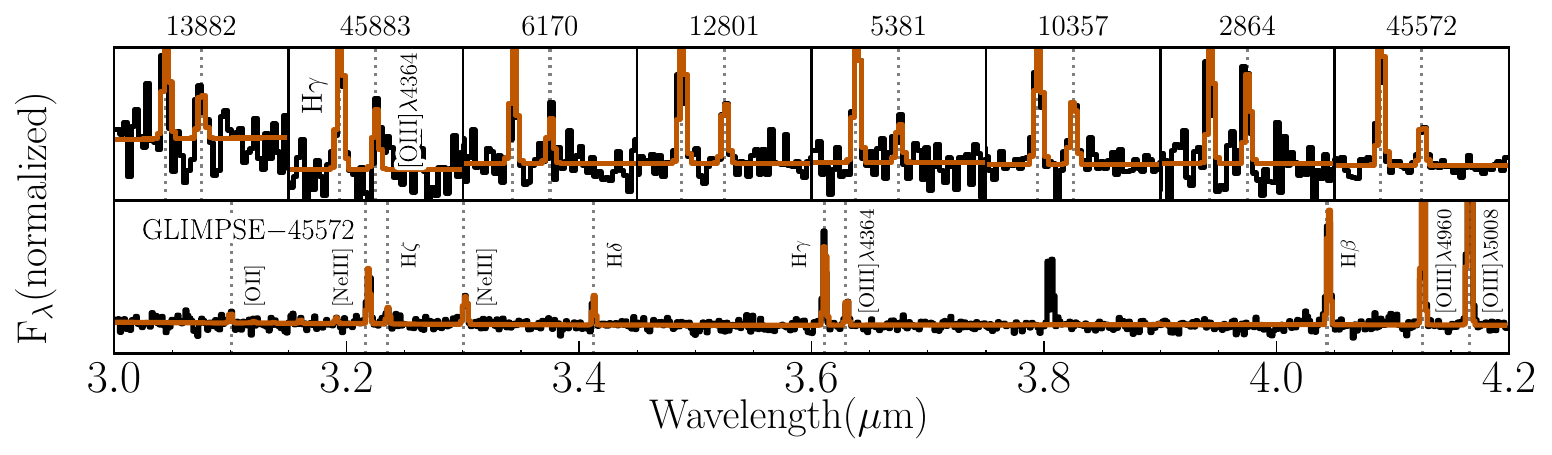}
\caption{GLIMPSE-D \OIIIwa\ emitter spectra shown in the black curves and the fits are shown in orange curves. The top panels show \OIIIwa\ as well as H$\gamma$ in each \OIIIwa\ emitter. The bottom panel illustrates a deep G395M spectrum of one of the \OIIIwa\ emitters: GLIMPSE-45572, demonstrating the power of GLIMPSE-D.
Emission lines are labeled with the dotted lines, including \OIIdw, \NeIIIw, H$\zeta$, \NeIIIwb, H$\delta$, H$\gamma$, \OIIIwa, H$\beta$ \OIIIwc, and \OIIIw.
}
\label{fig:line}
\end{figure*}

We remove potential interlopers and keep in the final sample only 8 targets consistent with being dominated by star formation.
GLIMPSE-D only obtained G395M spectroscopy, which covers \OIIIwa\ at $z\sim5.58-10.8$, where all of our samples are at $z\sim6-8$.
To increase the sample size and improve constraints on the $z\sim6-8$ MZR, we collect \OIIIwa\ emitters from the literature (\S\ref{sec:archival}) and reanalyze them using our homogeneous direct method.
Combined with our 8 GLIMPSE-D galaxies, this provides a more robust determination of the MZR.
We gather in total 13 \OIIIwa\ emitters with SNR(\OIIIwa) $>3$ at $z\sim6-8$ from literature.
The IDs, fitted spectroscopic redshifts, RA, DEC, and the reference of 8 \OIIIwa\ emitters from GLIMPSE-D and 13 \OIIIwa\ emitters from literature (``The $z\sim6-8$ Auroral Sample'' hereafter) is organized in Table \ref{tab:info}.
Figure \ref{fig:line} shows the line fits to \OIIIwa\ (and H$\gamma$) of 8 GLIMPSE-D samples, as well as a full G395M spectrum of GLIMPSE-45572.

\section{SFR, Mass, and Metallicity measured}
\label{Sec:analysis}

\subsection{SED fitting} \label{sec:SED}
\begin{figure}
\centering
\includegraphics[width=\columnwidth]{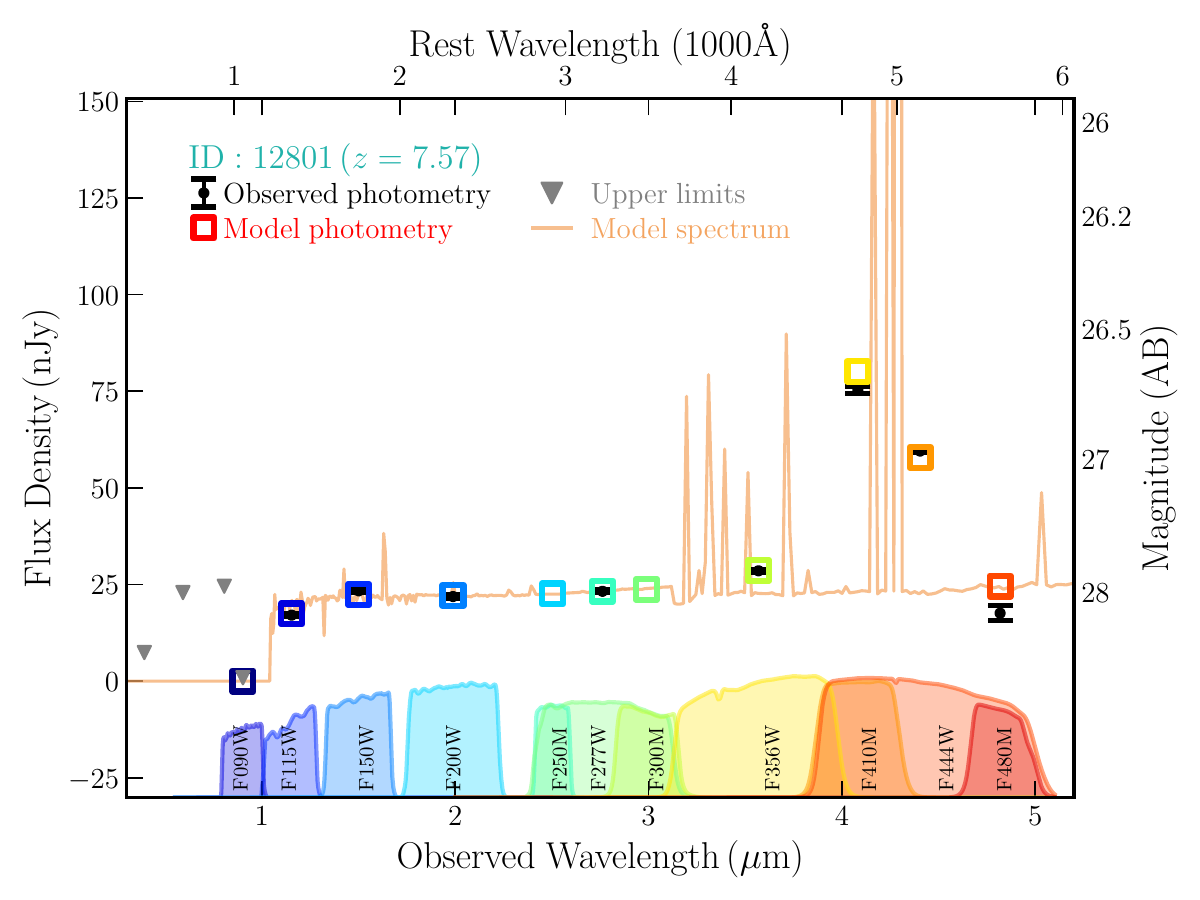}
\caption{
A \texttt{Bagpipes} SED fitting to GLIMPSE-12801. Colorful squares are model photometry while the black points are observed photometry. The orange curve shows the \texttt{Bagpipes} fit and the gray triangles are the upper limits where are undetected photometry.
}
\label{fig:SED}
\end{figure}

We perform SED fitting for all of the $z\sim6-8$ Auroral Sample on their photometry, using \textsc{Bagpipes} \citep{Carnall2018} to estimate their physical properties, including stellar mass and star-formation rates. 
The methodology, including adopted priors, assumptions, and model choices, is detailed in \citet{Hsiao2023}, unless stated otherwise.
In short, we adopt the Binary Population and Spectral Synthesis (BPASS) v2.2.1 models \citep{Stanway2018} for the inclusion of binary stellar evolution.
\textsc{Bagpipes} reprocessed nebular emission by the stellar templates with the photoionization code Cloudy \citep{Ferland2017}.
For the star formation history (SFH), we adopt a delayed-$\tau$ SFH with the \citet{Calzetti2000} dust attenuation law (with $0<A_{V}<8$ and a fixed $\eta=2$), and a \citet{Kroupa2001} initial mass function. 
The stellar metallicity varies between 0.5\% Z$_{\odot}$ to 200\% Z$_{\odot}$ as a prior, and we also fix the stellar metallicity to the gas-phase metallicity estimated using the direct metallicity measured, but the results did not change significantly.
The physical properties estimated from the SED fitting, including stellar masses and SFR are presented in Table \ref{tab:prop_SED}. 
An example SED fit to GLIMPSE-12801 is shown in Figure \ref{fig:SED}.

\subsection{Lensing Magnification}

We adopt a strong gravitational lensing model to correct the estimated stellar mass and SFR (see \S\ref{Sec:analysis}).
The details of the model can be found in \citet{Atek2025} and \citet{Furtak2026}, which is built upon the \texttt{Zitrin-analytic} model \citep{Zitrin2015}.
Briefly, the galaxy cluster Abell1063 is modeled with two components: the dark matter distribution and the galaxy distribution.
For the dark matter, a smooth component is parameterized as pseudo-isothermal elliptical mass distributions \citep{Kassiola1993}.
The member galaxies in the cluster are modeled as dual pseudo-isothermal ellipsoids \citep{Eliasdottir2007}.
We include two dark matter halos in Abell1063.
We utilize both spectroscopic data (spectroscopic redshifts) and imaging data (multiple lensed images) to constrain the model.

For the literature samples, we simply obtain the magnification of objects from the literature.
The magnifications of the $z\sim6-8$ Auroral Sample are presented in Table \ref{tab:prop_SED}.

\subsection{Dust Reddening Correction}
\label{sec:dust}
Before measuring the metallicity, we take the dust attenuation into account to correct the emission lines. 
To determine the dust reddening effect, we evaluate the color excess $E(B-V)$, adopting the dust attenuation law from \citet{Calzetti2000}. 
We derive $E(B-V)$ for each object from the observed Balmer decrement H$\alpha$/H$\beta$ and H$\gamma$/H$\beta$ (when no H$\alpha$), comparing the observed ratio to the intrinsic ratio of 2.75 and 0.475, respectively, predicted by Case B recombination under $T_e=20,000\,$K and $n_e=100\,{\rm cm^{-3}}$ \citep[][]{Osterbrock1989,Storey1995}.
For objects where the observed Balmer decrement implies no significant reddening (consistent with Case B within 1$\sigma$), we set $E(B-V) = 0$ and apply no dust correction.
Once $E(B-V)$ is determined, we use the Calzetti attenuation curve to deredden all emission line fluxes before proceeding with measuring the electron temperature and metallicity, and the uncertainty from the dust correction is propagated to other calculations later on.
We summarize the measured $E(B-V)$ of the $z\sim6-8$ Auroral Sample in Table \ref{tab:prop_spec}.

\subsection{Star-Formation Rates} \label{sec:SFR}
In addition to the SFRs determined using SED fitting, we also estimate the SFRs from Balmer lines (e.g., H$\alpha$ and H$\beta$), which are sensitive to the past 10 Myr of star forming activity.
We first estimate the H$\alpha$ intrinsic luminosity (dust-corrected) of each source.
Instead of the traditional \citet{Kennicutt1998} SFR conversion, we adopt an SFR conversion factor of $3.2 \times$ \tentotheminus{42}, which is more suitable for high-redshift galaxies with lower metallicity  \citep[$0.28\,Z_{\odot}$; ][]{Reddy2018}.
We then estimate the SFR assuming zero Lyman continuum escape fraction and a \citet{Chabrier2003} IMF; the results are presented in Table \ref{tab:prop_spec}.
We acknowledge that this conversion may overestimate SFRs for some of our galaxies, which have even lower metallicities ($\lesssim 0.1\,Z_{\odot}$; see \S\ref{sec:metallicity}).
At such low metallicities, enhanced ionizing photon production per unit SFR can lead to systematically higher H$\alpha$ luminosities, potentially overestimating our SFRs by $\sim0.2-0.3$ dex \citep{Kramarenko2026}.

\subsection{Metallicity Measurements} \label{sec:metallicity}

To determine the direct ($T_{e}$-based) oxygen abundances,
we use \pyneb\ \citep{Luridiana2015} with the collision strengths from \citet{Aggarwal1999} to calculate the corresponding oxygen abundances and physical conditions ($T_e$).
The total oxygen abundance can be approximated by the sum of $\rm{O^{++}}/\rm{H^{+}}$ and $\rm{O^{+}}/\rm{H^{+}}$, assuming higher excitation states as well as neutral oxygen are negligible:
\begin{equation}
    \rm{\frac{O}{H}}\simeq\frac{\rm{O^{+}}}{\rm{H^{+}}}+\frac{\rm{O^{++}}}{\rm{H^{+}}},
\end{equation}
where the ionic abundances can be obtained from:
\begin{equation}
    \frac{\rm O^{++}}{\rm H^{+}}=\frac{I_{\rm [OIII]}}{I_{\rm H\beta}}\frac{j_{\rm H\beta}}{j_{\rm [OIII]}},
\end{equation}
where $I$ is the emission line fluxes, and $j$ is the emissivity.
To determine the emissivity, the electron density ($n_e$) and $T_e$ are required.
\subsubsection{Electron Densities}
First, to assess the $n_e$, we note that GLIMPSE-D used the medium-resolution grating G395M, whose spectral resolution is insufficient to resolve density-sensitive doublets.
Since the observations of GLIMPSE-D were carried out only with G395M, the electron density doublets such as \OIIdw\ or \CIIIdw\ are not covered in most of the GLIMPSE-D \OIIIwa\ emitters at $6<z<8$.
Even if they are, the spectral resolution of $R\sim1000$ in G395M is not enough to resolve \OIIdw\ and thus measure the electron density.
In fact, \SIIdw\ is covered in most of the GLIMPSE-D \OIIIwa\ emitters, potentially offer electron density measurements, especially with the wider wavelength separation.
However, due to the high  degree of ionization and possibly low Sulfur abundance, we do not detect any of \SIIdw\ within our sample .

Recent JWST studies have shown that $n_{e}$ from different ionization zones can vary by as much as two orders of magnitude \citep[e.g.,][]{Mingozzi2022,Topping2024,Martinez2025}.
The O$^{++}$/H$^{+}$ determined from \OIIIw, and thus O/H, is strongly affected by high densities approaching the critical density of \OIIIw, $n_e \gtrsim10^5\,{\rm cm^{-3}}$ \citep[e.g.,][]{Hayes2025, Martinez2025,Arellano2026}.
Hence, it is important to study the possible density distribution of the gas, especially at high redshifts, where galaxies may have more extreme conditions \citep{Isobe2023,Abdurrouf2024}.

We therefore consider several assumptions:  (1) for our fiducial configuration, we adopt the redshift-dependent electron densities, $n_{e}=54\times(1+z)^{1.2}\,{\rm cm^{-3}}$, derived by \citet{Abdurrouf2024}, which fitted the redshift from $z=0$ out to $z\sim10$ with resolved \OIIdw\ assuming a uniform density model; (2) we assume constant electron densities, including $n_{e}=10^2, 10^3, 10^4,$ and $10^5\,{\rm cm^{-3}}$ to account for the possibility that extremely high densities can bias chemical abundance measurements \citep[e.g.,][]{Hayes2025,Martinez2025,Arellano2026}.

\subsubsection{Electron Temperatures and Metallicities}
\label{sec:ne}
Then, we determine the electron temperatures ($T_{\rm e}$(\OIII)) with detected \OIIIw\ and the auroral line \OIIIwa, using the \pyneb\ task \texttt{getTemDen}.
To estimate the low-ionization zone gas temperature (O$^+$), we apply the relation $T_{\rm e}$(\OII) = 0.7 $\times$ $T_{\rm e}$(\OIII) + 3000\,K \citep{Campbell1986}.
We also test different $T_{\rm e}$(\OII)-$T_{\rm e}$(\OIII) relations from \citet{Izotov2006} suitable for metal-poor galaxies, and find consistent metallicity.
With the electron temperatures and electron densities, we thus calculate the emissivity and populate the abundances, with the task \texttt{getIonAbundance}.

\begin{figure}
\centering
\includegraphics[width=\columnwidth]{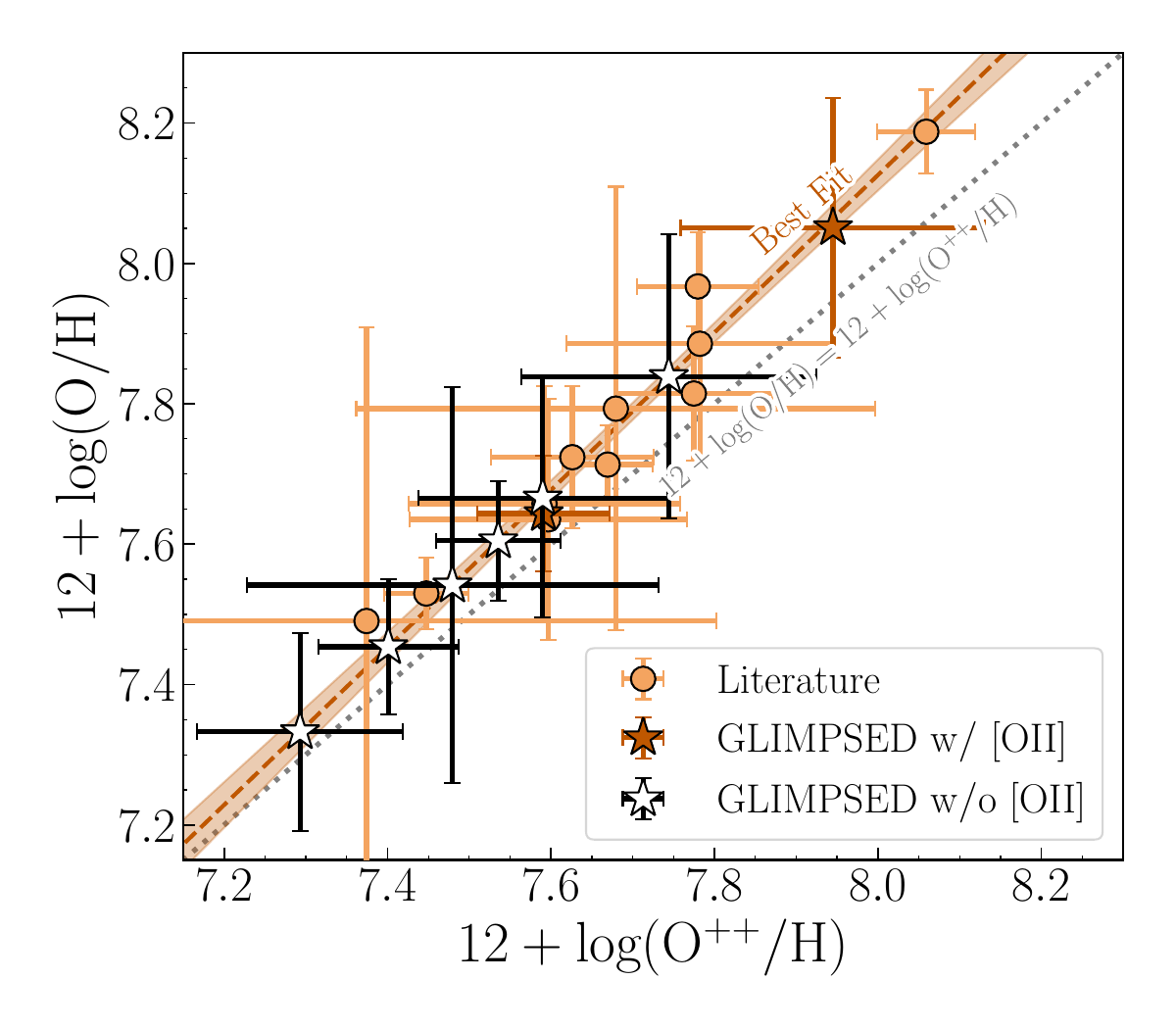}
\caption{12+log(O/H) vs 12+log(O$^{++}$/H) as a pseudo ICF. The literature (the light orange circles) and GLIMPSE-D \OIIIwa\ emitters with detected \OIIdw\ (the dark orange stars) are used to fit the relation presented in the orange dashed line.
12+log(O/H) of GLIMPSE-D \OIIIwa\ emitters without detected \OIIdw\ are obtained adopting the orange relations, demonstrated in the white stars.
The gray dotted line highlights the relation of 12+log(O/H) = 12+log(O$^{++}$/H).
This plot demonstrates that at lower metallicity the contribution from singly ionized oxygen is negligible.
}
\label{fig:ICF}
\end{figure}

However, given the wavelength coverage of 2.87--5.10$\,{\rm \mu m}$ for G395M, \OIIdw\ is only redshifted to the observable coverage at $z>6.7$.
Therefore, we only detect \OIIdw\ in two out of the eight \OIIIwa\ emitters in GLIMPSE-D.
To carefully determine the O$^+$ species, we first calculate the 12+log(O/H) and 12+log(O$^{++}$/H) for the 13 galaxies with \OIIdw\ detections in the $z\sim6-8$ Auroral Sample.
We fit the relation between 12+log(O/H) and 12+log(O$^{++}$/H) as a pseudo Ionization Correction Factor (ICF) where O$^+$ are unseen:
\begin{equation}
\begin{array}{l}
\label{eq:ICF}
    12+\log(\mathrm{O/H}) = (1.12\pm0.06) \\
    \qquad \times\left(12+\log(\mathrm{O^{++}/H})\right)-(0.85\pm0.47).
\end{array}
\end{equation}
Figure \ref{fig:ICF} shows the measured 12+log(O$^{++}$/H) and 12+log(O/H) with the fitted relation.
At $z\sim6-8$, the singly ionized species (estimated from \OIIdw) do not significantly change the total oxygen abundance.
\OIIIwa\ emitters with \OIIdw\ detections in the sample, show consistent abundance (still slightly higher) between 12+log(O/H) and 12+log(O$^{++}$/H).
We also find that galaxies with the lower 12+log(O$^{++}$/H) tend to exhibit negligible 12+log(O$^{+}$/H), and 12+log(O$^{++}$/H) can be used as a proxy of 12+log(O/H).
$T_e$, 12+log(O$^{+}$/H), 12+log(O$^{++}$/H), and 12+log(O/H) are presented in Table \ref{tab:prop_spec}.

\begin{deluxetable}{lccc}
\tablecaption{\label{tab:prop_SED}
The physical properties estimated using SED fitting from \texttt{Bagpipes} and the magnification.}
\tablewidth{\columnwidth}
\tablehead{
\colhead{ID} &
\colhead{Stellar Mass$^{a,b}$} &
\colhead{SFR$_{\rm SED}^{a,b}$} &
\colhead{$\mu^{c}$} 
\\
\colhead{} &
\colhead{log($M/M_\odot$)} &
\colhead{$M_\odot/{\rm yr}$} &
\colhead{} 
}
\startdata
\multicolumn{4}{c}{GLIMPSED} \\
\hline
13882 & $7.99^{+0.05}_{-0.05}$ & $5.16^{+0.51}_{-0.51}$ & $1.39$ \\
45883 & $6.73^{+0.02}_{-0.02}$ & $0.54^{+0.03}_{-0.02}$ & $1.38$ \\
6170 & $7.60^{+0.03}_{-0.03}$ & $4.10^{+0.27}_{-0.26}$ & $1.23$ \\
12801 & $7.83^{+0.05}_{-0.05}$ & $6.05^{+0.46}_{-0.49}$ & $1.31$ \\
5381 & $8.15^{+0.04}_{-0.08}$ & $5.86^{+0.81}_{-0.84}$ & $1.22$ \\
10357 & $7.66^{+0.06}_{-0.06}$ & $4.56^{+0.63}_{-0.61}$ & $1.29$ \\
2864 & $7.10^{+0.12}_{-0.14}$ & $0.53^{+0.04}_{-0.03}$ & $1.28$ \\
45572 & $7.22^{+0.02}_{-0.03}$ & $1.66^{+0.09}_{-0.10}$ & $1.46$ \\
\hline
\multicolumn{4}{c}{Literature} \\
\hline
GTO1199-2 & \ldots & \ldots & \ldots\\
COSMOS-419213 & \ldots & \ldots & \ldots\\
CEERS-01027 & $7.96^{+0.11}_{-0.07}$ & $9.27^{+2.78}_{-1.30}$ & \ldots\\
ERO-05144 & \ldots & \ldots & 3.18 \\
ERO-06355 & \ldots & \ldots & 1.78 \\
GLASS-10021 & $8.64^{+0.14}_{-0.33}$ & $23.57^{+9.17}_{-6.31}$ & 1.72 \\
ERO-10612 & \ldots & \ldots & 1.86 \\
EXCELS-48659 & $8.35^{+0.24}_{-0.21}$ & $7.82^{+1.59}_{-1.40}$ & \ldots\\
EXCELS-60713 & $8.12^{+0.18}_{-0.17}$ & $6.38^{+2.59}_{-1.46}$ & \ldots\\
GLASS-100003 & $8.24^{+0.11}_{-0.08}$ & $17.51^{+5.22}_{-2.96}$ & 1.34 \\
GOODSN-100026 & $7.81^{+0.22}_{-0.14}$ & $6.55^{+2.64}_{-1.85}$ & \ldots\\
GOODSN-100163 & $7.73^{+0.10}_{-0.07}$ & $5.35^{+1.36}_{-0.80}$ & \ldots\\
EXCELS-119504 & $8.22^{+0.29}_{-0.33}$ & $6.68^{+1.79}_{-1.50}$ & \ldots\\
\enddata
\tablenotetext{a}{Galaxies without mass/SFR measurements are lacking the requisite photometry in the archive.}
\tablenotetext{b}{Properties are delensed.}
\tablenotetext{c}{Magnification.}
\end{deluxetable}

\begin{deluxetable*}{lcccccc}
\tablecaption{\label{tab:prop_spec}
The dust attenuation, SFRs, fiducial physical conditions, and metallicities determined from NIRSpec spectroscopic data, assuming $n_e=54\times(1+z)^{1.2}\,{\rm cm^{-3}}$ for both GLIMPSED and literature \OIIIwa\ emitters.
}
\tablewidth{\columnwidth}
\tablehead{
\colhead{ID} &
\colhead{$E(B-V)$} &
\colhead{SFR$_{\rm H\alpha}^{a}$} &
\colhead{$T_e$(O$^{++}$)} &
\colhead{12+log(O$^{+}$/H)} &
\colhead{12+log(O$^{++}$/H)} &
\colhead{12+log(O/H)}
\\
\colhead{} &
\colhead{mag} &
\colhead{$M_\odot/{\rm yr}$} &
\colhead{K} &
\colhead{} &
\colhead{} &
\colhead{} 
}
\startdata
\multicolumn{7}{c}{GLIMPSED} \\
\hline
13882 & $0.18\pm0.06$ & $3.74^{+0.11}_{-0.11}$ & $ 21000\pm  6000$ & \ldots & $7.48\pm0.25$ & $7.54\pm0.28$ \\
45883 & $0.00$ & $1.64^{+0.05}_{-0.05}$ & $ 19000\pm  3000$ & \ldots & $7.59\pm0.15$ & $7.67\pm0.17$ \\
6170 & $0.00$ & $2.07^{+0.08}_{-0.08}$ & $ 18000\pm  3000$ & \ldots & $7.74\pm0.18$ & $7.84\pm0.20$ \\
12801 & $0.00$ & $3.52^{+0.14}_{-0.14}$ & $ 20000\pm  2000$ & $6.70\pm0.10$ & $7.59\pm0.08$ & $7.64\pm0.08$ \\
5381 & $0.00$ & $5.63^{+0.25}_{-0.25}$ & $ 14000\pm  2000$ & $7.39\pm0.18$ & $7.95\pm0.19$ & $8.05\pm0.18$ \\
10357 & $0.00$ & $4.29^{+0.09}_{-0.09}$ & $ 21000\pm  2000$ & \ldots & $7.54\pm0.08$ & $7.60\pm0.09$ \\
2864 & $0.00$ & $2.04^{+0.07}_{-0.07}$ & $ 24000\pm  4000$ & \ldots & $7.29\pm0.13$ & $7.33\pm0.14$ \\
45572 & $0.00$ & $3.37^{+0.12}_{-0.12}$ & $ 22000\pm  2000$ & $0.00$ & $7.40\pm0.09$ & $7.45\pm0.10$ \\
\hline
\multicolumn{7}{c}{Literature} \\
\hline
GTO1199-2 & $0.14\pm0.06$ & $104.32\pm3.07$ & $ 22000\pm  1000$ & $6.77\pm0.05$ & $7.45\pm0.05$ & $7.53\pm0.05$ \\
COSMOS-419213 & \ldots & \ldots & \ldots & \ldots & \ldots & \ldots \\
CEERS-01027 & $0.00$ & $9.69\pm0.37$ & $ 17000\pm  2000$ & $6.76\pm0.12$ & $7.77\pm0.10$ & $7.81\pm0.10$ \\
ERO-05144 & $0.00$ & $2.51\pm0.06$ & $ 16000\pm  1000$ & $7.51\pm0.09$ & $7.78\pm0.07$ & $7.97\pm0.08$ \\
ERO-06355 & $0.00$ & $16.00\pm0.37$ & $ 13000\pm  1000$ & $7.60\pm0.06$ & $8.06\pm0.06$ & $8.19\pm0.06$ \\
GLASS-10021 & $0.46\pm0.22$ & $128.12\pm54.86$ & $ 24000\pm 20000$ & $6.86\pm0.54$ & $7.37\pm0.43$ & $7.49\pm0.42$ \\
ERO-10612 & $0.00$ & $4.28\pm0.13$ & $ 19000\pm  1000$ & $6.70\pm0.07$ & $7.67\pm0.06$ & $7.71\pm0.06$ \\
EXCELS-48659 & $0.16\pm0.12$ & $6.40\pm0.60$ & $ 45000\pm 29000$ & \ldots & \ldots & \ldots \\
EXCELS-60713 & $0.00$ & $3.95\pm0.12$ & $ 18000\pm  5000$ & $7.16\pm0.32$ & $7.68\pm0.32$ & $7.79\pm0.32$ \\
GLASS-100003 & $0.00$ & $8.45\pm0.66$ & $ 20000\pm  3000$ & $6.57\pm0.21$ & $7.60\pm0.17$ & $7.64\pm0.17$ \\
GOODSN-100026 & $0.00$ & $4.18\pm0.29$ & $ 19000\pm  2000$ & $7.03\pm0.11$ & $7.63\pm0.10$ & $7.72\pm0.10$ \\
GOODSN-100163 & $0.12\pm0.06$ & $5.32\pm0.25$ & $ 17000\pm  3000$ & $7.21\pm0.18$ & $7.78\pm0.16$ & $7.89\pm0.16$ \\
EXCELS-119504 & $0.00$ & $6.08\pm0.46$ & $ 20000\pm  3000$ & $6.80\pm0.20$ & $7.59\pm0.17$ & $7.66\pm0.17$ \\
\enddata
\tablenotetext{a}{Properties are delensed.}
\end{deluxetable*}

\subsection{Strong-Line Diagnostics}
\label{sec:strong_line}

\begin{figure*}
\centering
\includegraphics[width=\textwidth]{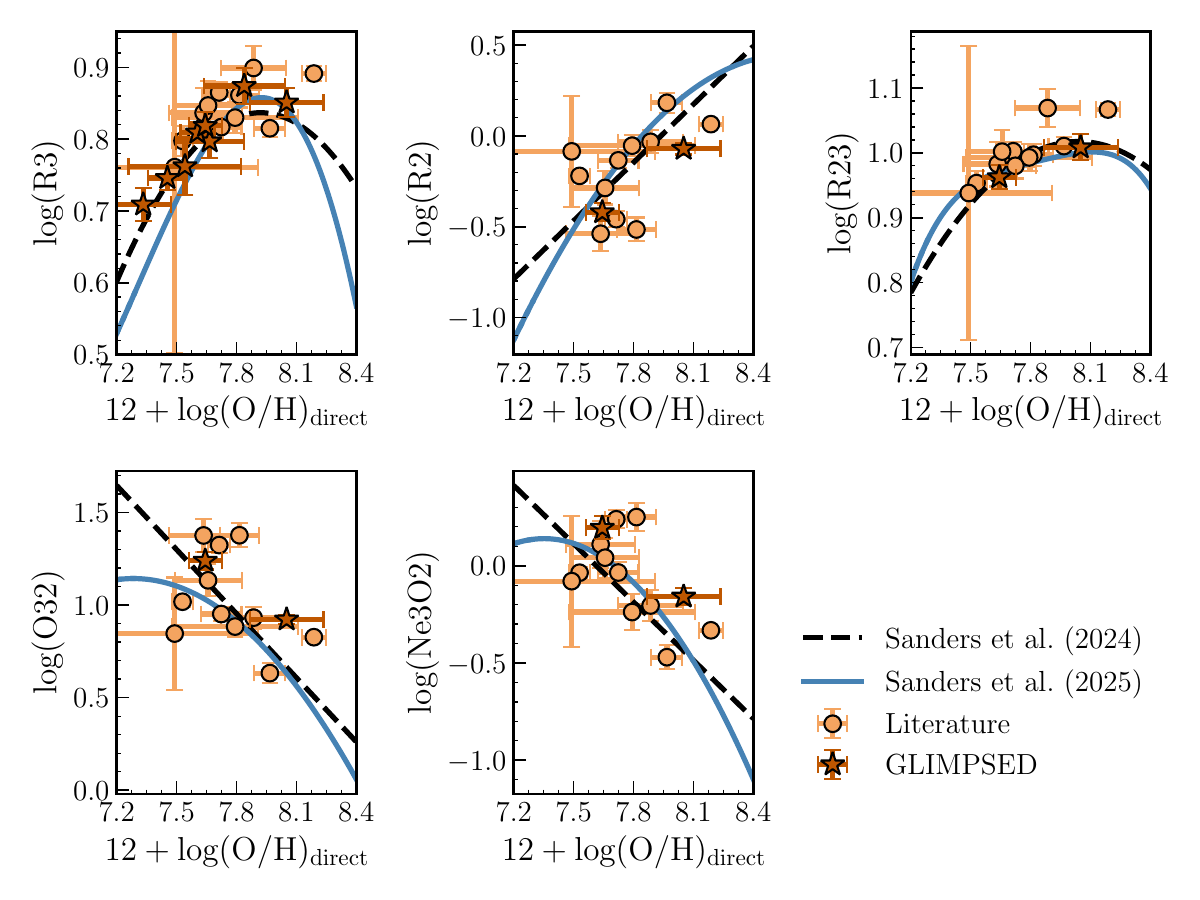}
\caption{Strong-Line diagnostics of the $z\sim6-8$ Auroral Sample at $6<z<8$.
The literature data are shown in the light-orange circles and the GLIMPSE-D \OIIIwa\ emitters are presented in the dark-orange stars.
For comparison, we show the strong-line diagnostics derived from a larger sample size spanning $z\sim2-9$ from \citet{Sanders2024} (the black dashed lines) and \citet{Sanders2025} (the blue lines), which are consistent with the shorter timeframe from the $z\sim6-8$ Auroral Sample at $6<z<8$.
}
\label{fig:diag}
\end{figure*}

We present the strong-line diagnostics in Figure \ref{fig:diag}, with the $z\sim6-8$ Auroral Sample as well as the strong-line relations from \citet{Sanders2024} and \citet{Sanders2025}, where they calibrated the strong-line relations from 30 and 139 galaxies, respectively, at $z\sim2-10$ with direct measurements with JWST.
Line ratios are defined as follows:

R3 = \OIIIw / H$\beta$,

R2 = \OIIdw / H$\beta$,

R23 =  (\OIIIdw + \OIIdw) / H$\beta$,

O32 = \OIIIw / \OIIdw, and

Ne3O2 = \NeIIIw / \OIIdw.

Overall, our strong-line diagnostics of \OIIIwa\ emitters agree well with the empirical relations calibrated from \citet{Sanders2024} and \citet{Sanders2025}.
They adopted larger samples across a wide redshift range of $z\sim2-10$.
This suggests that at a smaller time period of $z\sim6-8$, \OIIIwa\ emitters still follow similar strong-line relations as $z\sim2-10$.
In addition, we expand our sample to the fainter, low-mass regime of $M_{*}<10^8\,M_{\odot}$.
Although these GLIMPSE-D \OIIIwa\ emitters do not show significantly lower metallicities, the R3 relation remains consistent in this lower-mass, but similar metallicity regime.

Other than R3, however, we note that most of the remaining strong-line diagnostics agree well with our results.
These samples still have relatively higher metallicities than the extremely metal-poor regime.
Although consistent, direct metallicity measurements in even more metal-poor galaxies would be crucial to calibrate the low-metallicity end of strong-line diagnostics.

\section{Results and Discussion}
\label{Sec:discussion}

\begin{figure}
\centering
\includegraphics[width=\columnwidth]{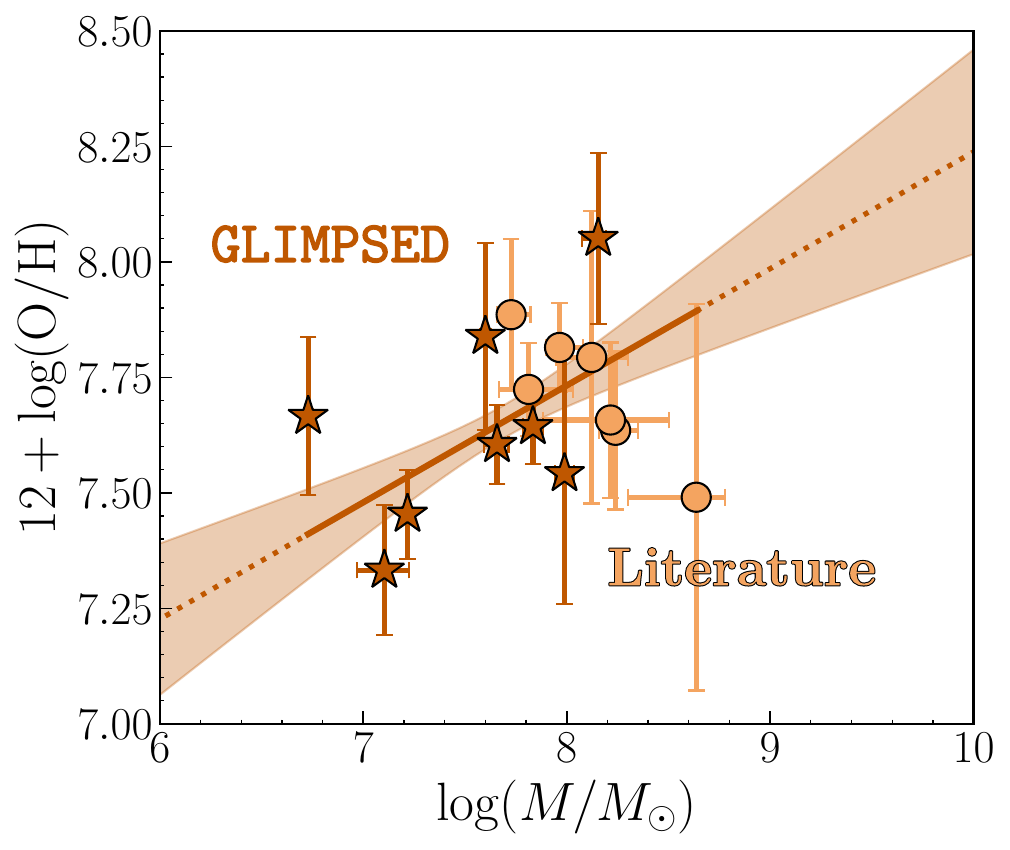}
\caption{Mass-metallicity relation using direct $T_e$ method of the $z\sim6-8$ Auroral Sample at $6<z<8$.
GLIMPSE-D \OIIIwa\ emitters are shown in the dark-orange stars while the literature \OIIIwa\ emitters are presented in the light-orange circles.
The linear fit to the MZR is shown in the orange solid line, and the orange dashed line indicates the extrapolation from the fit.
The orange shaded region highlights the $1\sigma$ uncertainty of the MZR fit.
The sample shows a positive mass-metallicity correlation at $z\sim6-8$, while with a large scatter, and the slope has a large uncertainty.
}
\label{fig:MZR}
\end{figure}

\subsection{Mass-Metallicity Relation at $6<z<8$}
\label{sec:MZR}

\begin{figure*}
\centering
\includegraphics[width=0.49\textwidth]{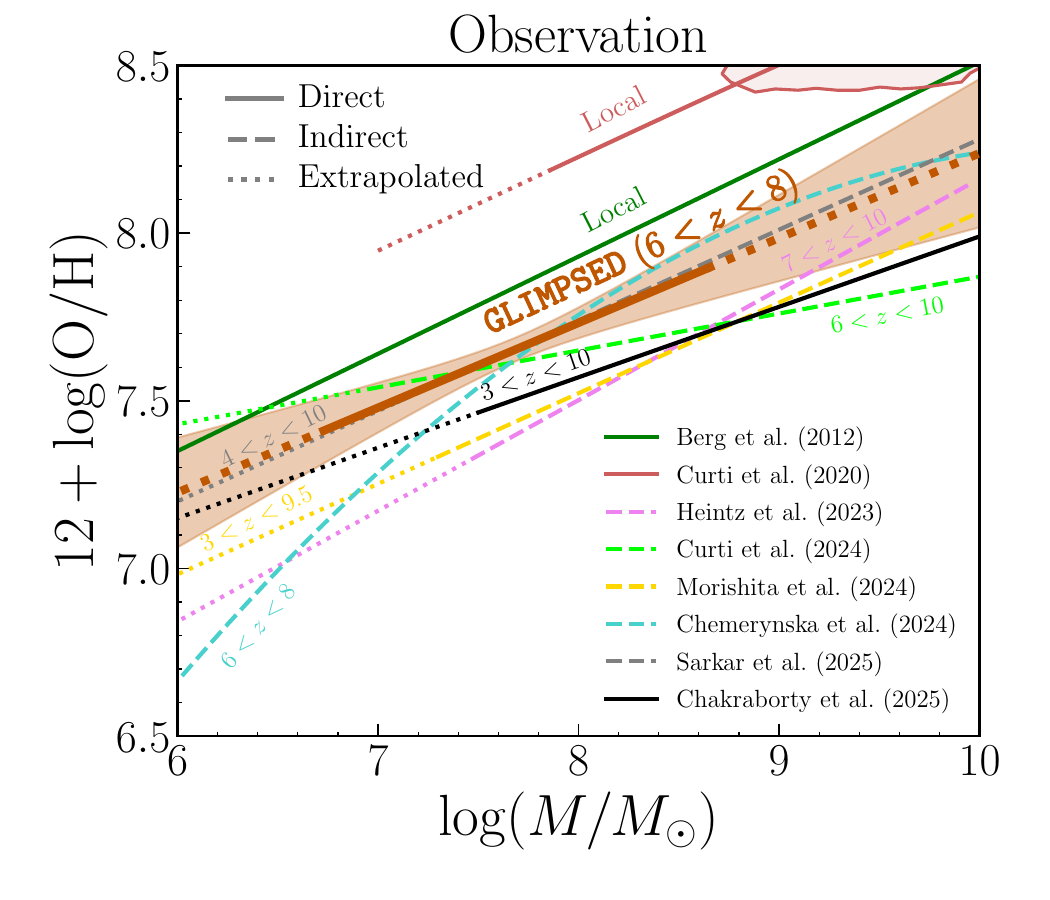}
\includegraphics[width=0.49\textwidth]{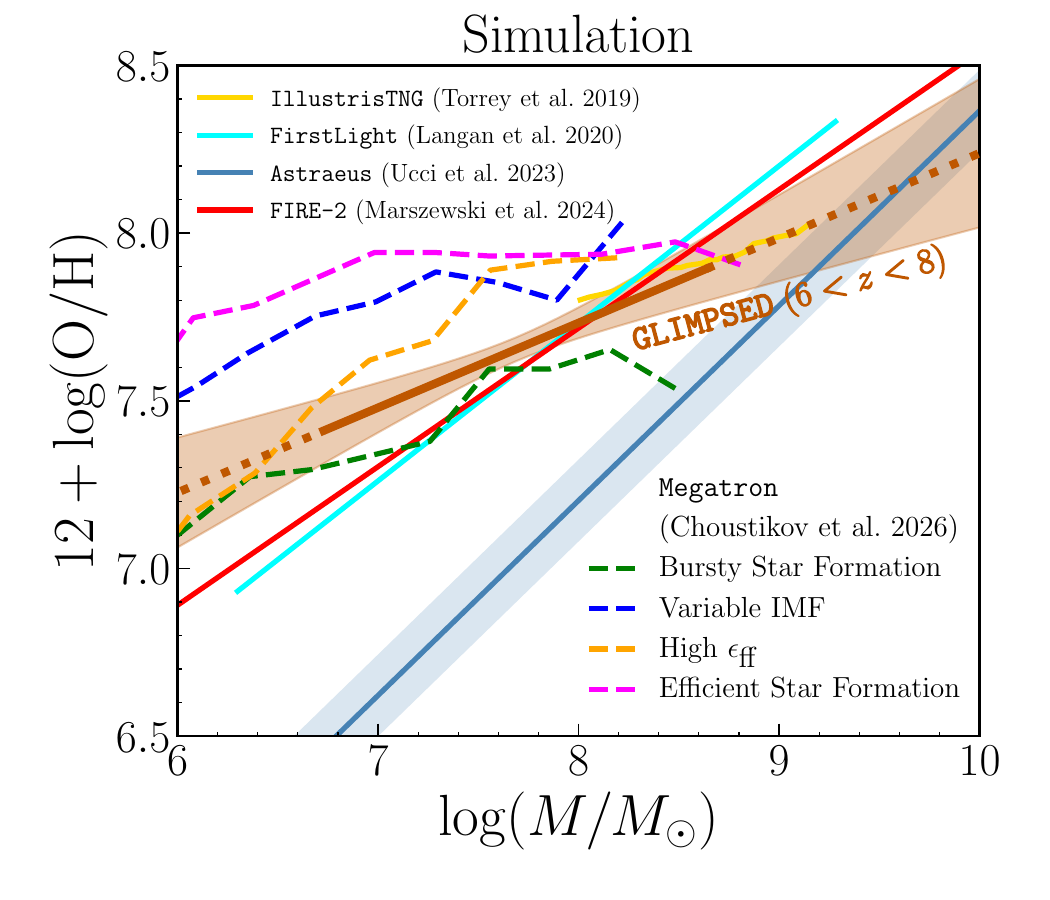}
\caption{Our fitted MZR at $z\sim6-8$ compared to observations (left panel) and simulations (right panel).
\textbf{Left panel}: We compared the MZR to the local MZR \citep[the green line;][]{Berg2012}, and other high-z observations: \citealt{Heintz2023} ($7<z<10$; the magenta line), \citealt{Curti2024} ($6<z<10$; the light-green line), \citealt{Morishita2024} ($3<z<9.5$; the yellow line), \citealt{Chemerynska2024} ($6<z<8$; the teal line), \citealt{Sarkar2025} ($4<z<10$; the grey line), and \citealt{Chakraborty2025} ($3<z<10$; the black line).
The relations from pure direct measurements (This work, \citet{Berg2012}, and \citet{Chakraborty2025}), are shown in the solid lines, and the indirect measurements involved relations are shown in dashed lines, where extrapolated regimes are highlighted in dotted lines. 
The GLIMPSE-D MZR is consistent with other high-z studies, but uniquely uses a purely direct method and extends measurements into the low-mass regime ($M_{*}<10^{8}\,M_{\odot}$).
\textbf{Right panel}: We compare the MZR to other high-z simulations: \texttt{IllustrisTNG} \citep{Torrey2019} in the yellow line, \texttt{FirstLight} \citep{Langan2020} in the cyan line, \texttt{Astraeus} \citep{Ucci2023} in the blue line, \texttt{FIRE-2} \citep{Marszewski2024} in the orange line, and \texttt{Megatron} shown in dashed lines \citep{Choustikov2025}.
Different configurations in \texttt{Megatron} are exhibited: Bursty Star Formation (green), Variable IMF (blue), High $\epsilon_{\rm{ff}}$ (orange), and Efficient Star Formation (magenta).
The observed GLIMPSE-D MZR is significantly flatter than most of the simulations but consistent with the slope predicted from \texttt{Megatron}.
}
\label{fig:MZR_comp}
\end{figure*}

\begin{figure*}
\centering
\includegraphics[width=\columnwidth]{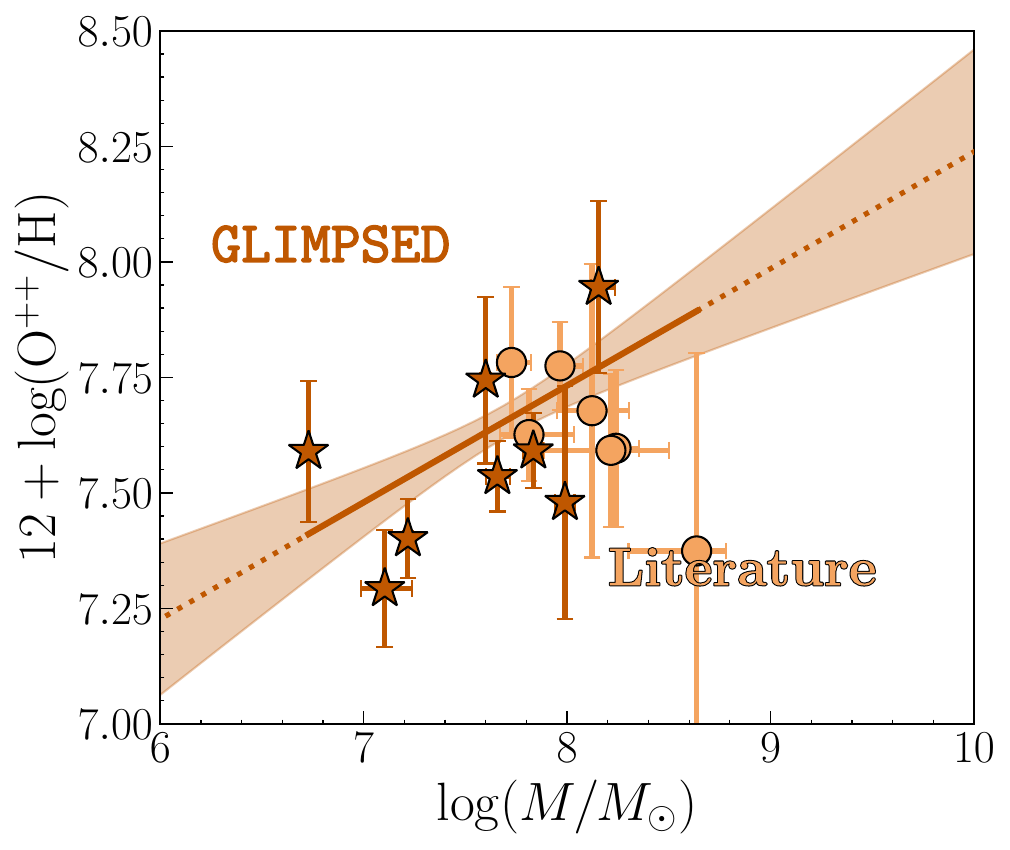}
\includegraphics[width=\columnwidth]{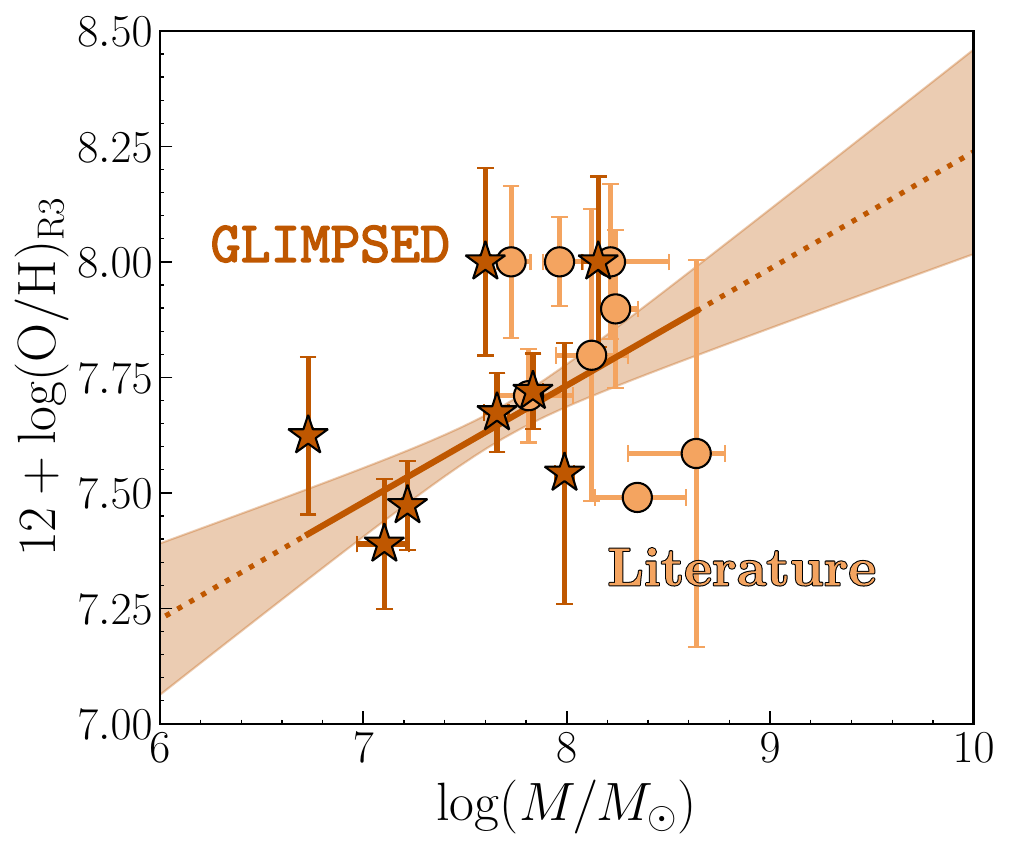}
\caption{
Variations of MZR, where the dark-orange stars show GLIMPSE-D data while the light-orange circles are literature data. The linear fits are the fit to the direct MZR in Figure \ref{fig:MZR}, which are shown in the orange solid lines, and the orange dashed lines indicate the extrapolation from the fits.
The orange shaded regions highlight the $1\sigma$ uncertainty of the MZR fit.
\textbf{Left panel}: The relation between stellar mass and doubly ionized oxygen abundance.
The MZR fit is steeper than the data, as expected from the pseudo-ICF in Figure \ref{fig:ICF}.
\textbf{Right panel}: MZR but the metallicity is estimated from the strong-R3 ratio instead of the direct $T_e$.
The strong-line MZR is consistent with the direct MZR.
}
\label{fig:other}
\end{figure*}

We present the GLIMPSE-D MZR, along with literature data at $z\sim6-8$, in Figure \ref{fig:MZR}.
We fit linear relations (in log-log space) to the MZR; the best-fit relation is:

\begin{equation}
\label{eq:MZR}
    12+{\rm log(O/H)}=(0.25\pm0.10)\,{\rm log}(M/M_{\odot})+(5.7\pm0.7).
\end{equation}
We compare the MZR with the local relation \citep{Berg2012,Curti2020}, other JWST high-z studies \citep{Heintz2023,Curti2024,Morishita2024,Chemerynska2024,Sarkar2025,Chakraborty2025,Lewis2025} in Figure \ref{fig:MZR_comp}.

The slope, normalization (overall metallicity level of the relation at fixed mass), and knee (or turnover mass; where the MZR begins to flatten with stellar mass) are all linked to distinct physical mechanisms and the processes that regulate galaxies (see the recent review by \citealt{Curti2025}). 
We emphasize at first that, given the limited sample sizes, most studies fit the MZR using the strong-line method, except \citet{Chakraborty2025}, who studied galaxies over a wider redshift range, $z\sim3-10$.
Second, the normalization, set by the equilibrium between metals produced by massive stars and those locked up in low-mass stars, reflects the nucleosynthetic stellar yields, as well as  inflows of pristine gas and outflows of metal-enriched gas.
Consistent with other JWST high-z MZR studies, our results indicate that galaxies in this early epoch ($z\sim6-8$) have lower metallicities than the local MZR at a given stellar mass. 
Our MZR is $\sim0.5$ dex more metal-poor than the local relation at intermediate stellar masses ($\sim10^{8-9}\,M_{\odot}$), and $\sim0.3$ dex more metal-poor in the low-mass regime ($<10^{8}\,M_{\odot}$).
This pattern might suggest that high-z galaxies have higher gas fractions compared to local galaxies \citep[e.g.,][]{Mannucci2010,Zahid2014,Marszewski2024}.
A continuous inflow of metal-poor gas dilutes the ISM metallicity.
For a fixed stellar mass, the higher gas reservoir at high-z means metals are distributed over a larger gas mass, directly lowering the observed metallicity, which also increases the sSFR (see more discussion in \S{\ref{sec:FMR}}).
At high-z, galaxies have had less time to undergo multiple cycles of star formation and chemical enrichment compared to local galaxies. 
This also leads to fewer cumulative enrichment episodes, and thus contributes to lower overall metallicities, in addition to dilution from higher gas fractions.

Second, feedback mechanisms associated with star-formation-driven outflows affect the slope of the MZR \citep{Tremonti2004,Peeples2011,Chisholm2018,Sanders2021,Curti2025}.
The slope reflects how the metal-loading factor, which describes the efficiency of outflows in removing metals relative to star formation locking them into stars, changes across the mass range.
The slope of our MZR, $0.25\pm0.10$, is consistent with some of the JWST high-z studies, including \citet{Heintz2023,Morishita2024,Sarkar2025,Chakraborty2025}, and comparable to the local MZR.
\citet{Curti2024} proposed that the slope could come from different wind driving modes, ranging from ``energy driven'' (local MZR has a slope of $\sim0.33$) to ``momentum driven'' (slope of $\sim0.17$) winds \citep{Dave2012,Guo2016}.
Additionally, if ISM densities are higher at high-z, a larger fraction of the energy injected by supernovae is radiated away through cooling, reducing the mass-loading factor and making feedback-driven outflows less efficient at removing gas and metals.
However, given a large uncertainty $\pm0.1$ dex in our slope fit, it is consistent with both feedback mechanism, and therefore it is difficult to draw a conclusion about the feedback mechanism here (see more discussion in \S\ref{sec:simu}).

Finally, no clear knee is observed in the $z\sim6-8$ Auroral MZR show in Figure \ref{fig:MZR}. 
It is possible that the knee lies in an uncharted mass range, i.e., $\gtrsim10^{10}\,M_{\odot}$
Therefore, the limited data do not yet allow us to draw firm conclusions.

Since we do not have \OIIw\ detections in most of the GLIMPSE-D \OIIIwa\ emitters (see \S\ref{sec:fitting}), many of their metallicities are scaled from samples with \OII\ detections.
Thus, to reduce possible bias, we show the relation between the doubly ionized oxygen abundance, 12+log(O$^{++}$/H), and stellar mass in Figure \ref{fig:other}.
As expected from O32 ratio and the pseudo ICF we correct in Equation \ref{eq:ICF}, where more metal-rich galaxies have more O$^{+}$.
Therefore, in more massive galaxies, 12+log(O$^{++}$/H) tends to be lower than the best-fit from the MZR in Equation \ref{eq:MZR}.
In addition, we show the MZR estimated from the strong line diagnostics (R3 = \OIIIw / H$\beta$; see \S\ref{sec:strong_line}) in Figure \ref{fig:other}.
Remarkably, we find no significant offset between the MZR derived from 
direct $T_e$ metallicities and that from R3-based strong-line calibrations (Figure \ref{fig:other}).
The R3 MZR exhibits similar scatter, slope, and normalization (see \S\ref{sec:strong_line}).
This agreement validates the use of R3 methods at $z\sim6-8$.
The consistency suggests that strong-line methods can reliably trace the MZR at high redshift, at least within our metallicity, mass, and redshift range.

It is unclear whether the MZR continues to evolve with redshift at $z>3$.
Early JWST studies reported mild evolution \citep[e.g.,][]{Nakajima2023,Curti2023}.
At $z\sim6-8$, we do not observe clear evolution.
However, the small sample size limits our ability to detect it.
In Figure \ref{fig:MZR_comp}, our $z\sim6-8$ sample remains offset relative to the local MZR, consistent with redshift evolution, and the slope, normalization, and the knee is similar to other JWST high redshift studies, including results based on indirect measurements. 
GLIMPSE-D galaxies probe the low mass end of the MZR, which helps constrain the relation at these redshifts more precisely.
This is the first time we can probe the low mass end of the MZR at such redshifts using direct metallicities, thanks to the exceptional depths of GLIMPSE and GLIMPSE-D.

\subsection{Impact of Electron Densities on the MZR}
\label{sec:ne_dis}
\begin{figure}
\centering
\includegraphics[width=\columnwidth]{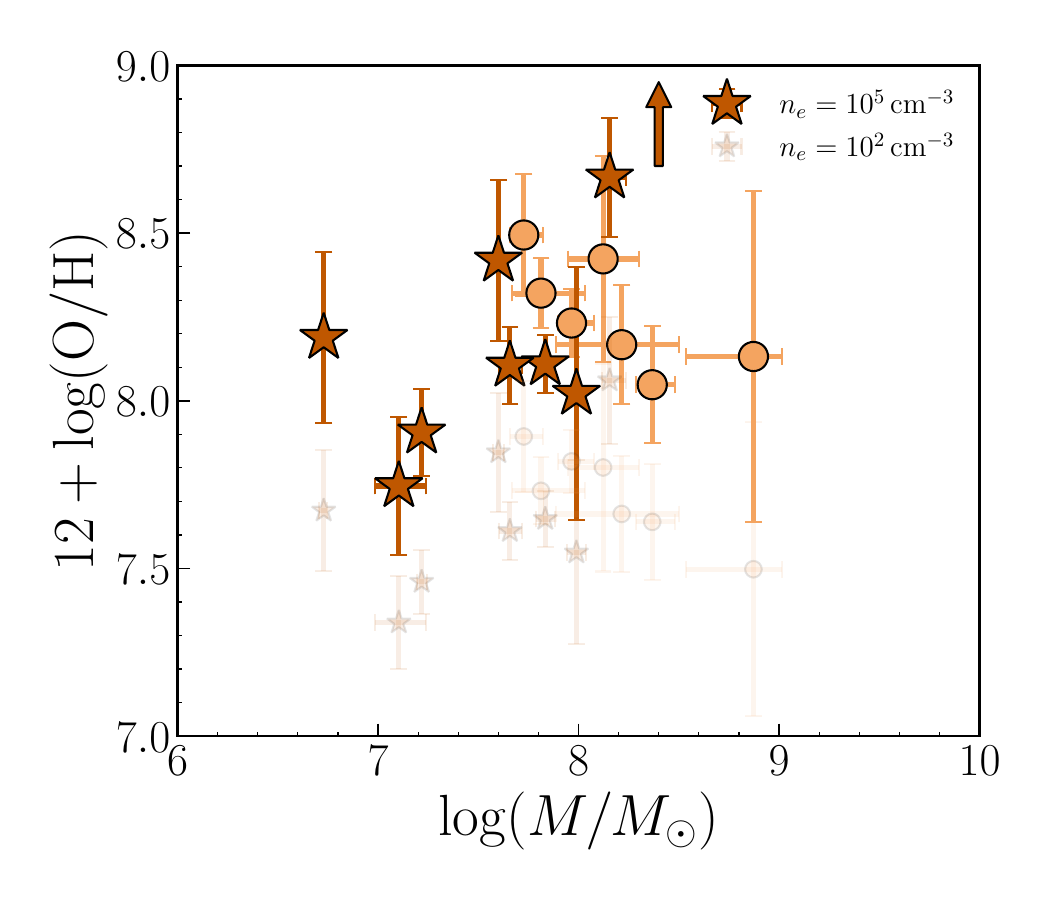}
\includegraphics[width=\columnwidth]{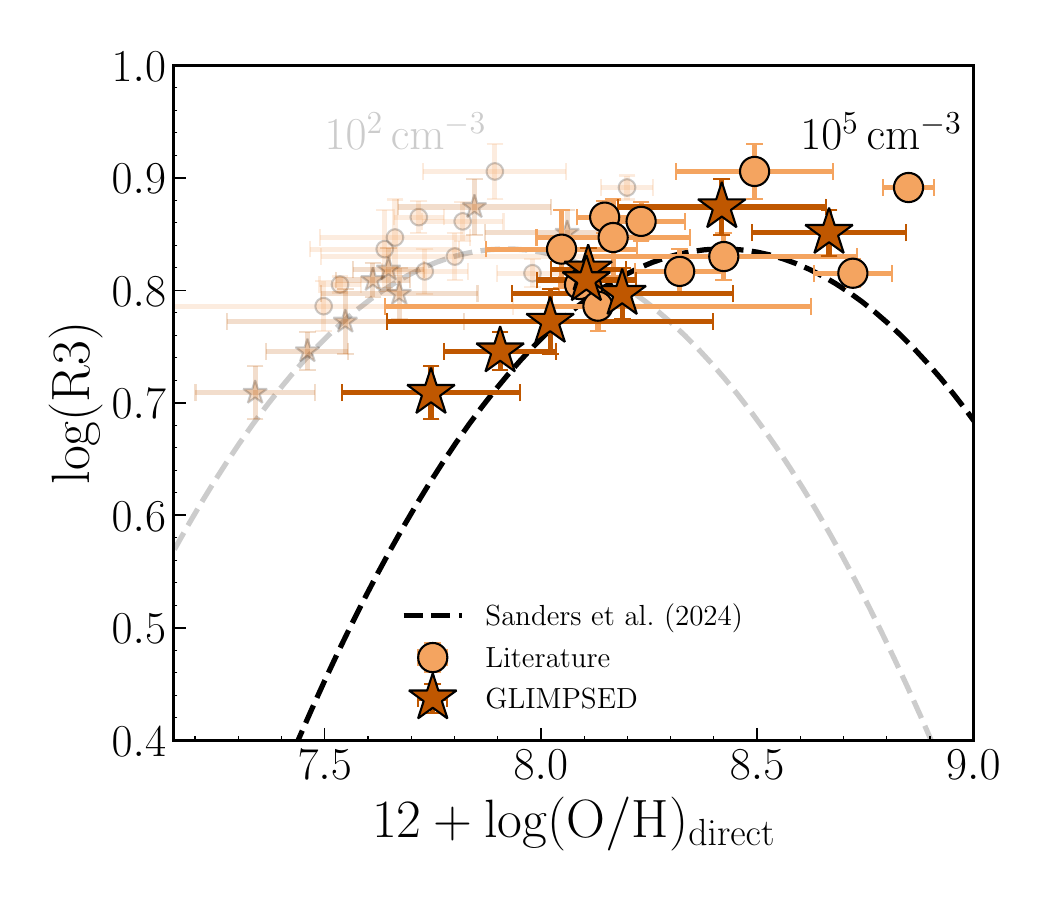}
\caption{\textbf{Top panel:} MZR under different electron density assumptions. $n_e=$ $10^2$ and $10^5\,{\rm cm^{-3}}$ are shown in brown stars (GLIMPSE-D), and orange circles (literature), with different transparency and sizes, from light to dark and small to large.
\textbf{Bottom panel:} Strong-line diagnostic of R3 under different electron density assumptions. Similar as top panel, $n_e=$ $10^2$ and $10^5\,{\rm cm^{-3}}$ are shown in brown stars (GLIMPSE-D), and orange circles (literature), with different transparency and sizes, from light to dark and small to large.
Again, we show the strong-line relation from \citet{Sanders2024}, and manually shift +0.5 dex in metallicity to match the extreme density condition.
Both panels demonstrate that metallicities will be highly underestimated by $\sim0.5$ dex if $n_{e}$ exceeds $\sim10^5\,{\rm cm^{-3}}$.
}
\label{fig:ne}
\end{figure}

As mentioned in \S\ref{sec:ne}, studies have shown electron densities can bias the metallicity measurements \citep{Hayes2025,Martinez2025}.
Motivated by these results, we assume a wide range of $n_e$, as well as adopting the redshift evolution relation, to test whether the electron densities play a critical role in MZR and affect our interpretation of chemical enrichment in the early universe.
In Figure \ref{fig:ne}, we show the metallicity measured under different assumed electron densities ($n_e=10^2$ and $10^5\,{\rm cm^{-3}}$) to assess the dependence of the MZR on density.
In the lower electron densities of $n_e=10^2, 10^3, 10^4\,{\rm cm^{-3}}$, the metallicities measured do not change significantly, and are consistent within $1\sigma$ error.
However, similar to the finding in the literature \citep[e.g.,][]{Hayes2025,Martinez2025}, in the case of extremely high electron densities of $n_e=10^5\,{\rm cm^{-3}}$, metallicities inferred increase rapidly and become $1-2\sigma$ higher than the low electron density assumptions, regardless of the stellar mass.
This occurs because, at such high densities, \OIIIw\ emission begins to be affected by collisional de‑excitation, reducing the \OIIIw\ line strength and biasing metallicity.
The extremely high-density conditions are expected, for instance, with massive star formation in compact HII regions \citep[e.g.,][]{Churchwell2002}, which might be the case for high-redshift metal-poor galaxies.

Despite the discrepancy of metallicity measurements, the overall slope of MZR remains similar, as shown in the top panel of Figure \ref{fig:ne}.
Again, we follow the identical process to derive the MZR fit.
The slopes of the relation between 12+log(O$^{++}$/H) and 12+log(O/H) are consistent with the fiducial configuration (assumed $n_e$ according to $z$), which are $0.25\pm0.10$ and $0.27\pm0.14$ for $n_e=10^2$ and $10^5\,{\rm cm^{-3}}$, respectively.
The variation in electron densities do not affect significantly the derived slopes, implicit that our conclusion earlier of different feedback mechanism and the bursty nature in the early universe remains valid (only if $n_e$ does not depend on stellar mass or metallicity).

In the bottom panel of Figure \ref{fig:ne}, we demonstrate how biased the strong line diagnostics (R3) will be if we do not treat the electron density correctly in high-density conditions.
If we assume a low electron density (seen in the local universe) but if the galaxies have an extremely high electron density, as mentioned above, metallicity will be highly underestimated.
In these high electron density conditions, the overall metallicities increase by up to $\sim0.7$ dex, similar to the finding of $\sim0.67$ dex from \citet{Martinez2025}.
The increasing metallicity is important for interpreting the normalization. 
Both the MZR and strong line diagnostics showcase the importance of measured electron densities at high redshift, where galaxies have more extreme conditions, in order to precisely determine the metallicity.
An expanded sample of electron density measurements  \citep[preferably in different ionization gases such as \OIIdw\ vs \CIIIdw; ][]{Berg2021,Harikane2025}, is needed at high redshift to further constrain the MZR, regardless of the method (direct-method vs strong-line diagnostics).

\subsection{MZR compared to simulations}
\label{sec:simu}

We compare the GLIMPSE-D MZR with the high-z simulations, including \texttt{IllustrisTNG} \citep{Torrey2019}, \texttt{FirstLight} \citep{Langan2020}, \texttt{Astaeus} \citep{Ucci2023}, \texttt{FIRE-2} \citep{Marszewski2024}, and \texttt{MEGATRON} \citep{Katz2025,Choustikov2025} in Figure \ref{fig:MZR_comp}.
Most of them predict a much steeper MZR at $z\sim6-8$ than we observe, including \texttt{FirstLight}, \texttt{Astaeus}, and \texttt{FIRE-2}.
Consistent with other early JWST studies, our results show a flatter MZR than these predictions, and early simulations have struggled to reproduce both the normalization and the slope at low stellar masses \citep{Curti2024}.

However, three of the four \texttt{Megatron} configurations \citep{Choustikov2025} predicted a flatter MZR even at the faint end, which matches our results.
These \texttt{Megatron} configurations tested different physics, including (1) Efficient Star Formation, (2) Bursty Star Formation, (3) Variable IMF, and (4) High $\epsilon_{\rm{ff}}$.
Efficient Star Formation is the fiducial run, with a feedback model tested to $z\sim0$ \citep{Agertz2021}.
However, as this model does not feature any boosted feedback, it ends up producing many more stars and ends up being more efficient on a global scale.
The Bursty Star Formation configuration has more bursty star formation histories because the feedback is stronger.
Variable IMF scenario allows the IMF to vary.
The High $\epsilon_{\rm{ff}}$ run features two changes.
(1) the local efficiency per free-fall time is set to 100\%, i.e. all gas in a given star-forming cell that can form stars will.
(2) The star-formation criteria have been changed in such a way as to preferentially form stars in lower-density regions.
This then means that emission lines are biased to these regions and any supernovae that go off will be better coupled to the gas as they explode into lower-density gas.
We refer readers to \citet{Katz2025} and \citet{Choustikov2025} for full details of each configuration and the simulation.

Except for the high $\epsilon_{\rm{ff}}$ configuration, the remaining three predict slopes similarly, and closer to the observed MZR from this work.
It is likely that in the high $\epsilon_{\rm{ff}}$ configuration, stars preferentially form in lower-density regions and therefore the stellar feedback is more coupled to the ISM to produce a steeper MZR.
On the other hand, given the fact that we do not see as  steep of a MZR as in the high $\epsilon_{\rm{ff}}$ case,  might indicate that stars still form in high-density regions, that the stellar feedback is less coupled to the ISM, and that stellar feedback is too inefficient to remove metals in the low-mass galaxies.

In the Bursty Star Formation scenario, stronger feedback reduces the overall normalization of the MZR by enhancing outflows and removing enriched gas, which affects both the mass and the abundance of a galaxy at later times.
Models with dense, turbulent gas, for example the efficient and bursty star formation models, can produce the flat slope observed at low stellar masses.
We caution that, given the limited sample size and the scatter in the MZR, firm conclusions remain difficult.

It has been suggested that high-redshift star formation may be burstier~\citep[e.g.,][]{Sun2023a,Sun2023b,Tacchella2023,Pallottini2023,Endsley2023,Endsley2025,Munoz2026}, which can flatten the slope of the MZR and increase its scatter~\citep{Pallotini2025}. 
However, in the \texttt{FIRE-2} simulation, which includes bursty star formation, the predicted slope remains much steeper than our results.
Additionally, across three configurations of \texttt{Megatron}, the inclusion of bursty star formation does not significantly change the slope, instead, the feedback prescriptions play the primary role.
Thus, we examine the scatter of our MZR compared with the local MZR.
To find the intrinsic scatter, we follow the methodology in \citet{Curti2024} and \citet{Chakraborty2025}, parameterized as:

\begin{equation}
    \sigma_{\rm MZR} = \sqrt{\sigma_{\rm obs}^2-\sigma_{\rm measured}^2},
\end{equation}
where $\sigma_{\rm obs}$ is the standard deviation of the $z\sim6-8$ Auroral Sample to the best-fit relation $\Bigl(\rm{std}\Bigl(\sum_{i=1}^{n}\bigl(Z_i - Z_{\rm fit}\bigr)\Bigr)\Bigr)
$ and $\sigma_{\rm measured}$ is the mean of the metallicity uncertainties $\Bigl(\frac{1}{n}\displaystyle\sum_{i}^{n}Z_{{\rm err},i}\Bigr)$.
We find the intrinsic scatter is not resolved ($\sigma_{\rm measured}^2>\sigma_{\rm obs}^2$), suggesting more accurate measurements are required (more data).
In addition, we also estimate the intrinsic scatter of the $z\sim6-8$ Auroral Sample but with the properties originally reported from the literature.
We find $\sigma_{\rm MZR}\sim0.2$ dex, higher than the scatter in the local MZR \citep[$\sim0.05-0.07$ dex;][]{Tremonti2004,Curti2020}.
We attribute the greater scatter to possible bursty star formation, where the regulation of galaxies in the early universe may be more chaotic.
More precisely, bursty star formation creates rapid, localized metal enrichment in galaxy centers, meaning metals cannot be efficiently redistributed across the galaxy before the next burst occurs, and therefore increases the scatter seen in the MZR \citep[e.g.,][]{Marszewski2025}.
In the \texttt{Megatron} simulations, although the Variable IMF configuration yields a slope similar to the Efficient Star Formation and Bursty Star Formation configurations, the Efficient and Bursty configurations exhibit much larger MZR scatter than the Variable IMF setup, which further supports that busty nature increase the scatter in MZR. 
However, with the limited data, as well as possible systematic errors, it is still impossible to conclusively confirm this.

One could argue that the flatter slope may originate from sample incompleteness, since more metal-rich or more massive galaxies are easier to detect and could bias the studied MZR.
Hence, we test the completeness.
We simply utilize the data from the \texttt{SPHINX} simulation \citep{Katz2023}, which is a high‑resolution cosmological radiation-hydrodynamics simulations that self‑consistently models early galaxy formation, it also releases the easily accessible data\footnote{\url{https://sphinx.univ-lyon1.fr}}.
We treat \texttt{SPHINX} as the ``true'' universe MZR, where \texttt{SPHINX} predicted a steeper MZR as well at $z\sim4-10$.
We adopt both GLIMPSE as well as GLIMPSE-D depths (ABmag $=30.82$ in F200W and $1.5\times10^{-20}\,{\rm erg\,s^{-1}\,cm^{-2}}$) to test if the unseen galaxies could bias the MZR, and filter which galaxies are detectable in brightness of both photometry and \OIIIwa.
However, we find that under the depth of GLIMPSE-D, if the slope is indeed steeper, we should be able to detect the more metal-poor objects (in the low-mass regime), and therefore constrain a steeper slope.
In other words, the flatter slope is less likely due to the fact of incompleteness.
Overall, the flat nature as well as the scatter, might both indicate that the regulation of mass and metallicity that we seen in the present day are not quite captured by current models.

\subsection{Fundamental Metallicity Relation}
\label{sec:FMR}

\begin{figure*}
\centering
\includegraphics[width=\columnwidth]{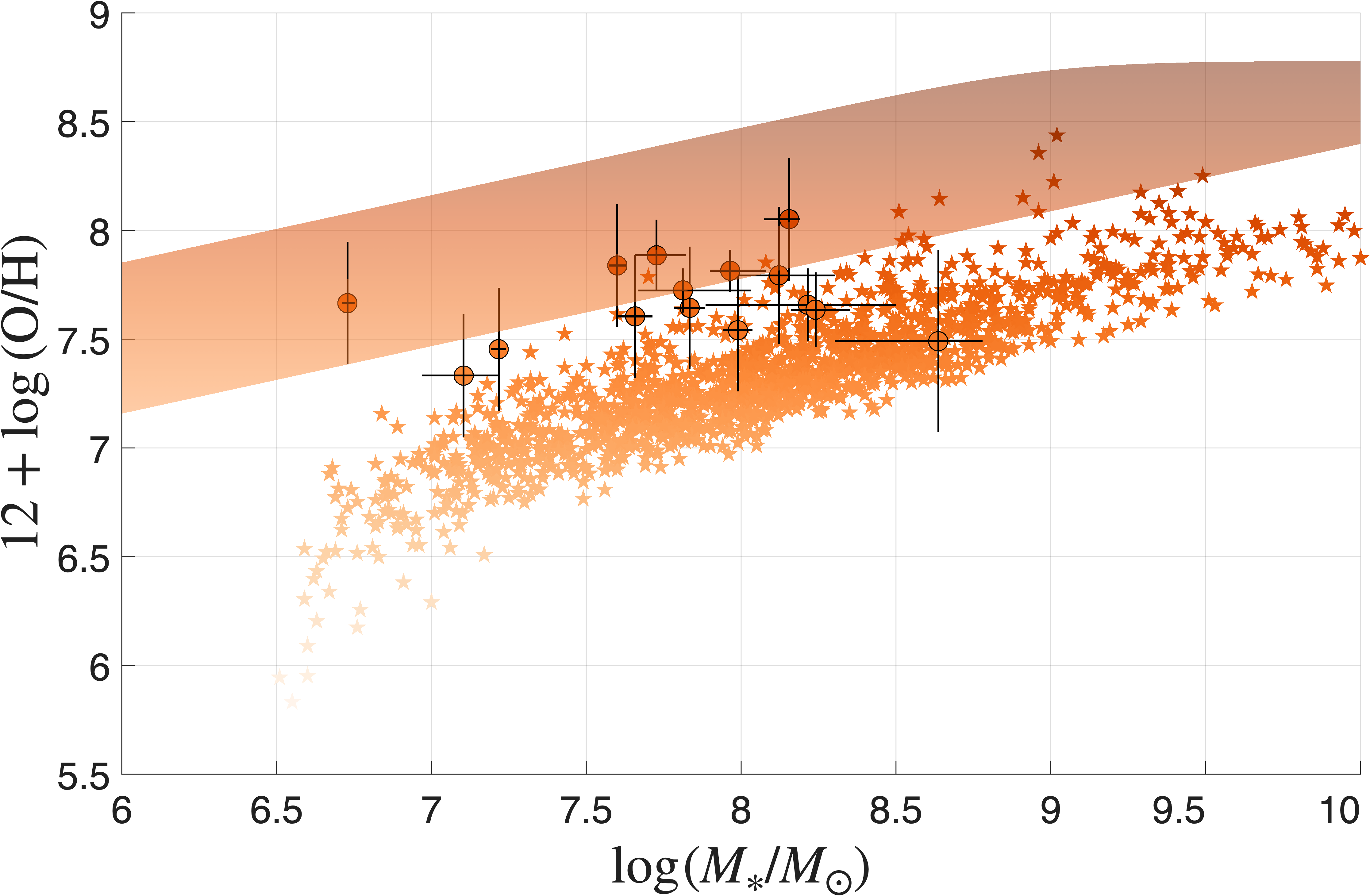}
\includegraphics[width=\columnwidth]{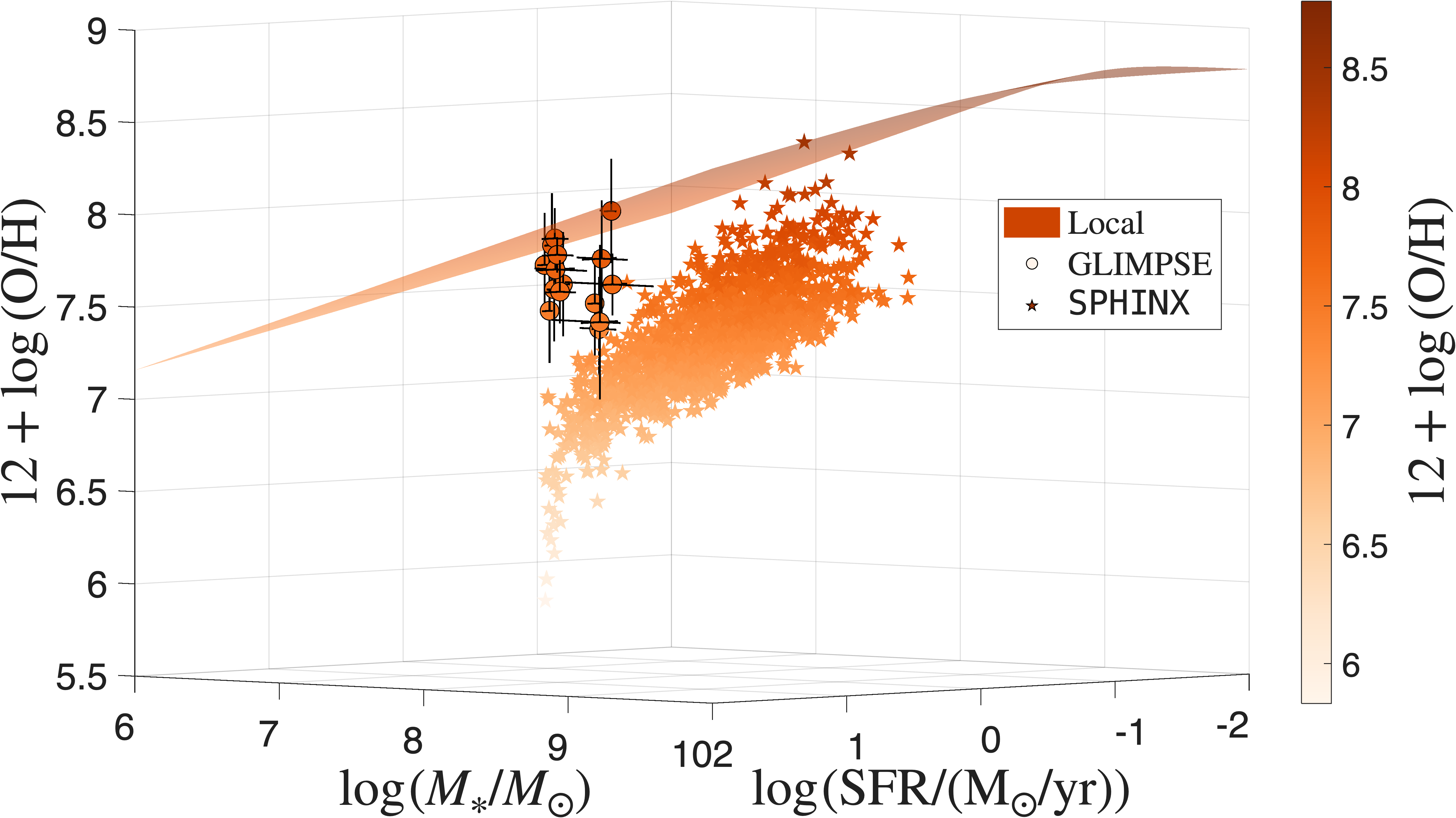}
\caption{3-dimensional fundamental metallicity relation shown in different angles. The gradient surface is the local FMR adopted from \citet{Curti2020}, and the data with errorbars shown are the $z\sim6-8$ Auroral Sample.
The gradient stars are galaxies in the high-z simulation: SPHINX \citep{Katz2023} at $z\sim4.6-10$.
For readers' convenience, we also provide an interactive 3D plot at
\url{https://mzr3d.s3.us-east-2.amazonaws.com/fmr3d_glimpse.html}.
}
\label{fig:FMR}
\end{figure*}

Here, we present, for the first time, an examination of the FMR using the direct $T_e$ method at $z\sim6-8$.
The three-dimensional FMR is shown in Figure \ref{fig:FMR}, with different viewing angles projected into two dimensions.
We plot the local FMR adopted from \citet{Curti2020}, parameterized as
\begin{equation}
    Z({\rm M,SFR})= Z_{0}-\gamma/\beta\,{\rm log}(1+(M/M_{0}({\rm SFR}))-\beta), 
\end{equation}
where ${\rm log(M_0 (SFR))} = m_0 + m_1\,{\rm log(SFR)}$, $Z_0= 8.779$, $m_0= 10.11$, $m_1= 0.56$, 
$\gamma = 0.31$, and $\beta  = 2.1$.
Similar to other JWST results, our FMR diverges from the local relation. 
Most of the $z\sim6-8$ Auroral Samples analyzed in this work fall below the FMR plane, indicating that, at fixed stellar mass and SFR, they are more metal-poor than galaxies in the nearby universe under the fiducial $n_e$ assumption.
On average, our galaxies are $\sim0.3$ dex below the local FMR, slightly smaller than the deviations found in (\citealt{Heintz2023}; $z\sim7-9$) and (\citealt{Curti2024}; $z>3$), and consistent with (\citealt{Nakajima2023}; $z\sim4-8$) and (\citealt{Laseter2025}; $z>3$).
We still clearly see a deviation from the local FMR at $z\sim6-8$.
Given our narrower redshift coverage and the smaller sample size of the $z\sim6-8$ Auroral Sample, we do not find significant redshift evolution.
We note that, even in the local universe, the low-mass regime of the FMR is not yet well studied.
For instance, \citet{Laseter2025} showed no clear FMR below $10^{9}\,M_{\odot}$, which may suggest that steady-state gas reservoirs are not yet established in these low-mass galaxies.

Interestingly, most of the $z\sim6-8$ Auroral Sample seem to follow a separate, vertical plane, as seen in the right panel of Figure \ref{fig:FMR}. 
In fact, this vertical plane is just the three dimensional expansion of the star forming main sequence \citep[e.g.,][]{Brinchmann2004,Speagle2014,Popesso2023}.
In the right panel of Figure \ref{fig:FMR}, we show the relation (two dimensional) of SFR and stellar mass, and the star forming main sequence can be clearly seen. 
This suggests that, in the three dimensional FMR at such redshift, the dependence of metallicity on SFR and stellar mass is weak compared to the nearby galaxies.
Furthermore, both theory and simulations have predicted that, in the early cosmic time, galaxies tend to accrete more pristine gas from the IGM \citep[e.g.,][]{Zahid2014,Marszewski2024}.
Another possible reason is selection bias, we only capture star forming galaxies, missing other quiescent and bursty galaxies.
But again, as described in \S\ref{sec:simu}, we check the completeness in the \texttt{SPHINX} simulation, where we should be able to see even burstier galaxies, although unknown selection bias in NIRSpec MOS target selection cannot be completely ruled out.

We compare the FMR to \texttt{SPHINX} \citep{Katz2023} simulation and the predictions at $z\sim4-10$ are shown in Figure \ref{fig:FMR}.
Interestingly, it predicts that galaxies should be much less enriched at $z\gtrsim4$.
Unlike \citet{Heintz2023}, who claimed that galaxies at $z\sim7-10$ are consistent with \texttt{Astraeus}, our FMR at $z\sim6-8$ is not consistent.
Similar to the finding in the two dimensional MZR shown in Figure \ref{fig:MZR_comp}, our results appear to be more metal rich than \texttt{Astraeus} at a given stellar mass and SFR, which again suggests the slope is flatter, with feedback less efficient at removing metals.
The scatter seen in the FMR indicates that metallicity is less tightly coupled to stellar mass and SFR, which can be attributed to bursty, stochastic star formation.
However, we note that simulations still struggle to predict a consistent FMR \citep[see][]{Garcia2025}.

As mentioned in \S\ref{sec:ne_dis}, metallicities are underestimated by $\sim0.5$ dex if lower $n_e$ values are assumed.
This $\sim0.5$ dex offset is comparable to the deviation between the local FMR and the $z\sim6-8$ Auroral Sample.
Accordingly, we investigate the FMR assuming $n_e=10^{5}\,{\rm cm^{-3}}$ for all galaxies.
The average metallicity of the $z\sim6-8$ Auroral Sample indeed increases toward the local FMR.
However, a distinct, vertically offset plane is still observed, though this could also be interpreted as scatter along the local FMR plane.
We caution that this scenario assumes a uniform $n_e=10^{5}\,{\rm cm^{-3}}$ for all galaxies, which is unlikely to be true; individual galaxies likely span a range of densities.
Robust $n_e$ measurements for individual objects are needed to conclusively test whether the FMR evolves beyond $z>3$.
If the metallicities of many high-z galaxies are indeed underestimated because their $n_e$ exceeds the critical density of \OIIIw, the apparent FMR evolution may be an artifact, and the FMR might not evolve at all.


\section{Conclusions}
\label{sec:conclusion}

In this paper, we present eight \OIIIwa\ emitters at $z\sim6-8$, using the GLIMPSE-D deep ($\sim30$ hrs) NIRSpec/G395M spectroscopic data. 
Owing to the unprecedented depth and gravitational lensing of GLIMPSE-D, these \OIIIwa\ emitters probe the low-mass regime (log$(M/M_{\odot})\sim6-8$) of the MZR.
We expand our sample by measuring the properties of archival 13 \OIIIwa\ emitters at the same redshift range in an uniform way.
With a total number of 21 \OIIIwa\ emitters, we establish the first MZR, as well as the FMR, purely from the most reliable direct $T_e$ based method, across a wide mass range of log$(M/M_{\odot})\sim6-10$ at $z\sim6-8$.
We summarize the main discoveries below:

\begin{enumerate}
    \item
    In agreement with other JWST high-z studies, our MZR has a $\sim0.3-0.5$~dex lower metallicity at fixed stellar mass than local relations.
    The metal-poor nature of high-z galaxies might be attributed to the higher gas fractions, which continuously inflow metal-poor gases to the ISM, making the overall metallicities lower.
    
    \item
    We fit the MZR to be $12+{\rm log(O/H)}=(0.25\pm0.10)\,{\rm log}(M/M_{\odot})+(5.7\pm0.7)$.
    The inferred MZR is consistent with \texttt{Megatron} simulations where high-redshift stars form in denser gas, leading to lower energy coupling of stellar feedback to the gas in the galaxy.
    However, the large uncertainty in the fitted slope hinders us in drawing a conclusive picture.

    \item We compare the strong-line diagnostics, specifically for the galaxies at $z\sim6-8$, which is the first diagnostics including the faint end at high-redshift.
    We do not find a significant deviation from the strong-line calibration published in the literature at a larger redshift range of $z\sim2-10$.

    \item We vary our assumed electron densities from $n_{e}=10^2-10^5\,{\rm cm^{-3}}$ and conclude that with low electron densities ($n_e<10^4\,{\rm cm^{-3}}$) as well as the varying with redshift ($n_{e}=54\times(1+z)^{1.2}\,{\rm cm^{-3}}$), the metallicity does not change significantly.
    However, in the high density assumption ($n_e\gtrsim10^{5}\,{\rm cm^{-3}}$), which near the critical density for \OIIIw, metallicity increases by $\sim0.5$ dex.
    Despite this, we find the overall slope of MZR does not change significantly, but only the normalization as long as $n_e$ does not correlate with $M_*$ or metallicity.
    However, the calibration of strong-line diagnostics would be broken under extremely dense regions if originally those direct metallicities came from the incorrect $n_e$, highlighting the necessity of high-z electron density measurements.

    \item We explore the FMR at $z\sim6-8$ from pure \OIIIwa\ emitters, and find that it is  not consistent with the local ($z\sim0-3.5$) FMR.
    Interestingly, we witness a new plane in the 3-D FMR.
    The plane is almost perpendicular to the local MZR.
    We suggest the dependence of metallicity on the SFR and stellar mass is much weaker than in nearby galaxies, and the plane is the mass-SFR relation of the star-forming main sequence, which might because early galaxies have accreted more pristine gas from IGM regardless of their stellar mass and SFR.  

This work highlights the power and the importance of deep spectroscopy, such as GLIMPSE-D, to detect the faint, low-mass galaxies to study the complete picture of MZR at high redshift.
We also demonstrate the importance of electron density.
Future observations, preferably with high-resolution grating onboard JWST, with larger surveys, and deeper to probe the electron density, deliver a larger sample, and lower mass, will be crucial to understand the chemical evolution and how galaxies evolved. 

\end{enumerate}

\section{Acknowledgments}
TH, VK, LJF, AA, and MGS thank the University of Texas at Austin Cosmic Frontier Center and Department of Astronomy for supporting this work.
This work is based on observations made with the NASA/ESA/CSA JWST, obtained at the Space Telescope Science Institute, which is operated by the Association of Universities for Research in Astronomy, Incorporated, under NASA contract NAS5-03127.
The JWST data presented in this article were obtained from the Mikulski Archive for Space Telescopes (MAST) at the Space Telescope Science Institute.
These observations are associated with programs GO \#3293 and DDT \#9223.
Support for program \#3293 and 9223 was provided by NASA through a grant from the Space Telescope Science Institute, which is operated by the Association of Universities for Research in Astronomy, Inc., under NASA contract NAS 5-03127.



%

\vspace{5mm}



\appendix


\bibliography{papers}{}
\bibliographystyle{aasjournal}



\end{document}